\documentclass[12pt]{article}
\usepackage[hmargin=1.75cm, vmargin=2.5cm]{geometry}
\usepackage{authblk}
\usepackage{graphicx}
\usepackage{natbib}
\usepackage{multirow}%
\usepackage{amsmath,amssymb,amsfonts}%
\usepackage{hyperref}
\usepackage{booktabs}%
\usepackage{bm}
\global\long\def\tr{\operatorname{tr}}
\global\long\def\SP{\operatorname{SP}}

\global\long\def\E{\operatorname{E}}
\global\long\def\Cov{\operatorname{Cov}}
\global\long\def\Var{\operatorname{Var}}
\global\long\def\ARE{\operatorname{ARE}}
\global\long\def\rank{\operatorname{rank}}

\global\long\def\iidsim{\stackrel{\text{i.i.d.}}{\sim}}
\global\long\def\calT{\mathcal{T}}

\global\long\def\calK{\mathcal{K}}

\global\long\def\diag{\operatorname{diag}}

\global\long\def\b#1{{\bf \bm{\mathit{#1}}}}
\global\long\def\0{\b 0}
\global\long\def\1{\b 1}
\global\long\def\bA{\b A}
\global\long\def\bC{\b C}
\global\long\def\bD{\b D}
\global\long\def\bM{\b M}

\global\long\def\bV{\b V}
\global\long\def\bb{\b b}

\global\long\def\bI{\b I}
\global\long\def\bJ{\b J}
\global\long\def\bn{\b n}
\global\long\def\bx{\b x}
\global\long\def\by{\b y}

\global\long\def\bB{\b B}
\global\long\def\bX{\b X}

\global\long\def\bE{\b E}
\global\long\def\bQ{\b Q}
\global\long\def\bP{\b P}

\global\long\def\bGamma{\b{\Gamma}}
\global\long\def\bOmega{\b{\Omega}}
\global\long\def\bSigma{\b{\Sigma}}
\global\long\def\bfeta{\b{\eta}}
\global\long\def\bphi{\b{\phi}}

\global\long\def\bfeta{\b{\eta}}
\global\long\def\h#1{\hat{#1}}
\global\long\def\hd{\h d}
\global\long\def\halpha{\h{\alpha}}

\global\long\def\hbSigma{\h{\bSigma}}
\global\long\def\hbGamma{\h{\bGamma}}
\global\long\def\hbOmega{\h{\bOmega}}
\global\long\def\hbM{\h {\bM}}
\global\long\def\hI{\h{I}}
\global\long\def\hT{\h{T}}
\global\long\def\t#1{\tilde{#1}}

\global\long\def\tbB{\t{\bB}}
\global\long\def\tbE{\t{\bE}}

\global\long\def\barby{\bar{\by}}
\global\long\def\barbx{\bar{\bx}}

\newcommand{\be}{\begin{equation}}
	\newcommand{\ee}{\end{equation}}

\newcommand{\T}{\top}
\newcommand{\B}{\text{\tiny{B}}}
\newcommand{\N}{\text{\tiny{N}}}
\newcommand{\MFW}{\text{\tiny{MFW}}}
\newcommand{\MFLH}{\text{\tiny{MFLH}}}
\newcommand{\MFP}{\text{\tiny{MFP}}}
\newcommand{\FLH}{\text{\tiny{FLH}}}
\newcommand{\ZZC}{\text{\tiny{ZZC}}}
\newcommand{\QCZ}{\text{\tiny{QCZ}}}
\newcommand{\QFZ}{\text{\tiny{QFZ}}}
\newcommand{\Z}{\text{\tiny{Z}}}
\renewcommand{\H}{\mbox{H}}

\newtheorem{theorem}{Theorem}
\newtheorem{remark}{Remark}

\newtheorem{example}{Example}
\newtheorem{lemma}{Lemma}
\newtheorem{proposition}{Proposition}

\begin{document}
	
	\title{\textbf{Modified  Tests of Linear Hypotheses Under  Heteroscedasticity for Multivariate Functional Data with Finite Sample Sizes}}
	\date{}
	
	\author[1]{\small Tianming Zhu}
	
	\affil[1]{\footnotesize{National Institute of Education, Nanyang Technological University, Singapore}}
	
	\maketitle

\begin{abstract}
As big data continues to grow, statistical inference for multivariate functional data (MFD) has become crucial. Although recent advancements have been made in testing the equality of mean functions, research on testing linear hypotheses for mean functions remains limited. Current methods primarily consist of permutation-based tests or asymptotic tests. However, permutation-based tests are known to be time-consuming, while asymptotic tests typically require larger sample sizes to maintain an accurate Type I error rate. This paper introduces three finite-sample tests that modify traditional MANOVA methods to tackle the general linear hypothesis testing problem for MFD. The test statistics rely on two symmetric, nonnegative-definite matrices,  approximated by Wishart distributions, with degrees of freedom estimated via a U-statistics-based method.  The proposed tests are affine-invariant, computationally more efficient than permutation-based tests, and better at controlling significance levels in small samples compared to asymptotic tests.  A real-data example further showcases their practical utility.

\vspace{0.5cm}
\end{abstract}

\section{Introduction}\label{intro.sec}
Over the past few decades, functional data have garnered significant attention due to their widespread collection across various scientific fields. In the early literature on functional data analysis (FDA), each functional observation was typically represented by a single curve, a framework referred to as the univariate case. This approach was sufficient when only one type of functional information was available for each individual. However, with rapid advancements in data recording and storage technology, researchers can now collect multiple streams of functional information for the same individual, such as measurements taken across different modalities or time points. As a result, interest in multivariate functional data (MFD) has grown significantly, where each observation is represented by multiple interrelated curves or functions.

Despite the growing interest in MFD, extending it from univariate data remains an emerging and underdeveloped field. The complexity increases as researchers must account for interactions between different functional components, such as dependencies across various measurement dimensions or functional processes. Consequently, new statistical models and techniques are required to adequately analyze and interpret MFD. This emerging framework holds significant promise for advancing our understanding of complex phenomena across disciplines; however, considerable work remains to develop a comprehensive theory and practical methods for MFD.

This study was partially inspired by the work of \cite{soh2023regularised}, which aimed to differentiate the geographical origins of extra virgin olive oil (EVOO) using attenuated total reflectance-Fourier-transform infrared (ATR-FTIR) spectra. In their study, 43 bottles of organic EVOO, spanning seven different brands from Greece, Spain, or Italy, were obtained from local suppliers. For each bottle, a portion of oil was extracted, stirred, and four drops were taken using a dropper. Each drop was scanned twice, resulting in a total of 344 spectra. The ATR-FTIR spectra were obtained using a Perkin-Elmer Spectrum 100 model with an ATR accessory, with spectral readings taken in the mid-infrared region, covering wavenumbers from 4000 $\mbox{cm}^{-1}$ to 500 $\mbox{cm}^{-1}$, at intervals of 1 $\mbox{cm}^{-1}$ (see \citealt{soh2023regularised} for details). The authors proposed the sparse fused group lasso logistic regression (SFGL-LR) model to distinguish between Greek and non-Greek EVOOs. However, \cite{soh2023regularised} noted that the spectral observations were not entirely independent, as each drop was scanned twice. Despite this, the model demonstrated good prediction accuracy and improved interpretability, especially with a large number of observations. In this context, introducing the concept of MFD is reasonable, where two scans of each drop represent a single functional observation for each origin. Given that geographical origin can influence consumer purchasing decisions, it is worth examining whether the mean spectral observations from these three regions differ.

The problem of comparing the mean vectors of $k$ multivariate populations based on $k$ independent samples is considered multivariate analysis of variance (MANOVA). In the context of functional data, when observations encompass more than one feature, the objective is to test the equality of the vector of mean functions across  $k$ independent functional populations. This is  referred to as one-way functional MANOVA (FMANOVA). \cite{gorecki2017multivariate} was the first to define the ``between" and ``within" matrices for MFD and used them to construct various test statistics. 
Building on this, \cite{Qiu2021} proposed two global tests for the two-sample problem in MFD, based on the integration and supremum of the pointwise Hotelling’s $T^2$-test statistic.  More recently, \cite{qiu2024tests} extended this work to the multi-sample problem. \cite{Zhu2021} explored the Lawley–Hotelling trace test for the FMANOVA problem under the assumption of equal covariance function matrices across the $k$ samples, while \cite{zhu2023ksampFMD} introduced a global test statistic tailored for the heteroscedastic FMANOVA problem. Additionally, \cite{zhu2023general} addressed the general linear hypothesis testing (GLHT) problem for MFD. However, 
the simulation results in Tables~\ref{size.tab1}, \ref{size.tab3}, and \ref{size.tab4} reveal that the aforementioned asymptotic tests perform poorly in terms of accuracy when the sample size is small, as they fail to maintain the desired significance level. Consequently, this study focuses on developing new tests for linear hypotheses under heteroscedasticity in MFD, particularly for cases involving finite sample sizes.

Throughout this paper, we write $\by(t)\sim \SP_p(\bfeta,\bGamma)$ for a $p$-dimensional stochastic process $\by(t)$ over a compact space $\calT$ with mean function $\bfeta:\calT\to\mathbb{R}^p$, covariance function $\bGamma:\calT^2\to\mathbb{R}^{p\times p}$, and $p\in\mathbb{N}$. At each time point $t\in\calT$, $\by(t)$ is a $p$-dimensional vector.  Hence, it is appropriate to employ transitional methods  from MANOVA in the context of MFD. 
Traditional one-way MANOVA tests for multivariate normal distributions rely on two independent Wishart matrices that share the same variance structure.  Examples include  the Wilks' likelihood ratio test \cite{wilks1932certain}, the Lawley–-Hotelling trace test \cite{lawley1938generalization,hotelling1951generalized}, the Bartlett–-Nanda–-Pillai trace test \cite{bartlett1939note,nanda1950distribution,pillai1955some}, and Roy's largest root test \cite{roy1953heuristic}. These tests are commonly used to test the equality of mean vectors across groups under the assumption of homogeneous covariance matrices. However, when this assumption is violated, the problem becomes more complex and is referred to as the $k$-sample Behrens–Fisher (BF) problem \cite{behrens1928beitrag,fisher1935fiducial}, or heteroscedastic one-way MANOVA. In such cases,  classical MANOVA tests may exhibit substantial bias, particularly when sample sizes are unbalanced across groups.  One approach to addressing the multivariate BF problem is the Wishart-approximation method,  first introduced by \cite{nel1986solution}.  This approach is conceptually similar to the $\chi^2$-approximation method developed by \cite{satterthwaite1946approximate}, where parameters are computed by matching the first two moments. \cite{harrar2012modified} was among the first to propose using a Wishart distribution to approximate these two Wishart matrices, building on the results of \cite{bathke2008nonparametric}. Later, \cite{xiao2016modified} noted that the modified tests by \cite{harrar2012modified} lacked affine invariance and extended their work, along with \cite{zhang2012approximate}, to address the GLHT problem in heteroscedastic two-way MANOVA. More recently, \cite{zeng2024robust} developed a robust test for multivariate repeated measures data. However, to the best of our knowledge, the application of the Wishart-approximation method to functional data has not yet been explored in the literature. 

In this paper, we propose three modified tests for linear hypotheses under heteroscedasticity in MFD with finite sample sizes, using the Wishart-approximation method. The key contributions of this work are as follows. First, we define the global variation matrices associated with the hypothesis and error for the GLHT problem and develop modified tests based on these matrices, with approximate degrees of freedom estimated from the data. Second, we demonstrate that the proposed tests are affine-invariant and consistent, regardless of the contrast matrix used to define the hypothesis. Third, we derive the approximate degrees of freedom using a U-statistics-based approach, which provides bias-reduced estimators. 
Simulation results in Section~\ref{Simu.sec} show that the proposed tests achieve similar size control as the permutation-based tests introduced by \cite{gorecki2017multivariate}, while outperforming these methods in terms of computational efficiency. Additionally, the proposed tests demonstrate better performance than those from \cite{Qiu2021, Zhu2021, zhu2023ksampFMD, zhu2023general, qiu2024tests} in maintaining the desired significance levels.

The rest of the paper is organized as follows.  The main results are presented in Section~\ref{Main.sec}. Simulation studies and real data applications are given in Sections~\ref{Simu.sec} and~\ref{real.sec}, respectively.  Concluding remarks are provided  in Section~\ref{con.sec}. Technical proofs of the main results are included  in Appendix.

\section{Main Results}\label{Main.sec}
\subsection{Models and Hypotheses}\label{model.sec}
Suppose we have $k$  independent multivariate functional samples given by $p$-dimensional stochastic processes:
\be\label{ksamp.sec2}
\by_{i1}(t),\ldots,\by_{in_i}(t)\iidsim \SP_p(\bm{\eta}_i,\bGamma_i),\; i=1,\ldots,k,
\ee
where $\bm{\eta}_i:\calT\to\mathbb{R}^p$ are the unknown vector of group mean functions of the $k$ samples and $\bGamma_i:\calT^2\to\mathbb{R}^{p\times p}$ are the unknown matrix of group covariance functions for all $i=1,\ldots,k$. In this paper, we study the following GLHT problem for MFD: 
\be\label{GLHTH0.sec2}
\H_0: \bC\bM(t)=\bC_0(t),\; t\in\calT \mbox{ vs } \H_1: \bC\bM(t)\neq \bC_0(t), \; \mbox{for some }t\in\calT,
\ee
where $\bC: q\times k$ is a known full-rank coefficient matrix with $\rank(\bC)=q\leq k$, and at time point $t\in\calT$, $\bM(t)=[\bm{\eta}_1(t),\ldots,\bm{\eta}_k(t)]^\T$ is a $k\times p$ matrix whose rows are the $k$ mean functions  and $\bC_0(t)$ is some known constant matrix.  It is worth noting that the GLHT problem (\ref{GLHTH0.sec2}) provides a broad and general framework that includes various specific hypotheses. To provide further clarity on the GLHT problem (\ref{GLHTH0.sec2}), we present the following examples of null hypotheses, all of which can be addressed within the framework of (\ref{GLHTH0.sec2}).
\begin{example}[Linear hypothesis testing with univariate functional data]
	
	For univariate functional data where $p=1$, the $k$ independent functional samples in (\ref{ksamp.sec2}) can be represented as $y_{i 1}(t),\ldots,$ $y_{in_i}(t)\iidsim \SP(\eta_{i},\gamma_{i}),\;i=1,\ldots,k$. In this scenario,  we have $\bM(t)=[\eta_1(t),\ldots,\eta_k(t)]^\T$ and the GLHT problem in (\ref{GLHTH0.sec2}) simplifies to the scenario  studied in \cite{smaga2019linear}. In particular, using the contrast matrix $\bC=(\bI_{k-1},-\bm{1}_{k-1})$,  where $\bI_k$ and $\bm{1}_k$ denote the identity matrix of size $k\times k$ and a $k$-dimensional vector of ones, respectively, and $\bC_0(t)=\bm{0}$ for all $t\in\calT$, the GLHT problem  (\ref{GLHTH0.sec2}) becomes equivalent to a one-way functional ANOVA (FANOVA).
\end{example}
\begin{example}[Two-sample test for MFD]
	For the special case where $k=2$,   the objective is to test whether the mean functions from two populations are equal, as $  \H_0: \bfeta_1(t)=\bfeta_2(t),\;  t\in\calT$ vs $\H_1: \bfeta_1(t)\neq \bfeta_2(t), \; \mbox{for some }t\in\calT$. 
	By using the configuration $\bC=(1,-1)$ and $\bC_0(t)=\bm{0}$ for all $t\in\calT$, the GLHT problem (\ref{GLHTH0.sec2}) simplifies to a two-sample problem, which has been addressed by \cite{Qiu2021}. 
\end{example}
\begin{example}[One-way FMANOVA]\label{fmanova.ex}
	In a broader context than the previous example, this involves testing the equality of mean functions across $k$ populations:
	\be\label{BFH0.sec2}
	\H_0: \bm{\eta}_1(t)=\cdots =\bm{\eta}_k(t),\;  t\in\calT,
	\ee
	against the usual alternative that at least two of the mean functions are not equal.  By using a contrast matrix $\bC=(\bI_{k-1},-\bm{1}_{k-1})$ and $\bC_0(t)=\bm{0}$ for all $t\in\calT$,  the problem becomes a one-way FMANOVA, as studied in \cite{gorecki2017multivariate, Zhu2021,zhu2023ksampFMD, qiu2024tests}.
\end{example}

\begin{example}[Testing linear combinations of mean functions] 
	The GLHT problem (\ref{GLHTH0.sec2}) can also encompass the testing of linear combinations of functional means as a special case, which can be described as $\H_0: \sum_{i=1}^k c_i \bfeta_i(t)=\bm{0},\;  t\in\calT$  vs $\H_1$: not $H_0$. The above problem can be equivalently tested by setting $\bC=(c_1,\ldots,c_k)$ and $\bC_0(t)=\bm{0}$ for all $t\in\calT$, and is primarily discussed in \cite{zhu2023general}.
\end{example}

\subsection{Test Statistics}
Based on the $k$ functional samples (\ref{ksamp.sec2}), the unbiased estimators of vector of mean functions $\bm{\eta}_i(t),i=1,\ldots,k$ and matrix of group covariance functions $\bGamma_i(s,t),i=1,\ldots,k$ are given by
\be\label{kmeancov.sec2}
\begin{array}{rcl}
	\hat{\bm{\eta}}_i(t)&=&\barby_i(t)=n_i^{-1}\sum_{j=1}^{n_i}\by_{ij}(t),\;i=1,\ldots,k, \;\mbox{ and }\\
	\hbGamma_i(s,t)&=&(n_i-1)^{-1}\sum_{j=1}^{n_i}[\by_{ij}(s)-\barby_i(s)][\by_{ij}(t)-\barby_i(t)]^\T,\; i=1,\ldots,k,
\end{array}
\ee
respectively. We begin by setting up the notation. 

Let $\hbM(t)=[\barby_1(t),\ldots,\barby_k(t)]^\T$ and $\bD_n=\diag(1/n_1,\ldots,1/n_k)$. To test hypothesis (\ref{GLHTH0.sec2}), we construct the pointwise variation matrix due to hypothesis for the heteroscedastic GLHT problem (\ref{GLHTH0.sec2}), denoted as $\bB_n(t)=[\bC\hbM(t)-\bC_0(t)]^\T(\bC\bD_n\bC^\T)^{-1}[\bC\hbM(t)-\bC_0(t)]$  for all $t\in\calT$. It follows that the global variation matrix due to hypothesis can be  defined as
\be\label{Bn.sec2}
\bB_n=\int_{\calT}[\bC\hbM(t)-\bC_0(t)]^\T(\bC\bD_n\bC^\T)^{-1}[\bC\hbM(t)-\bC_0(t)]dt.
\ee
Under the null hypothesis in (\ref{GLHTH0.sec2}), we can express $\bB_n$ in (\ref{Bn.sec2}) as 
\[
\bB_{n,0}=\int_{\calT}[\bC\hbM(t)- \bC\bM(t)]^\T(\bC\bD_n\bC^\T)^{-1}[\bC\hbM(t)- \bC\bM(t)]dt=\int_{\calT}\bX(t)^\T\bm{H}_n\bX(t)dt,
\]
where $\bX(t)=[\barbx_1(t),\ldots,\barbx_k(t)]^\T$ and 
$\bm{H}_n=\bC^\T(\bC\bD_n\bC^\T)^{-1}\bC=(h_{ij}):k\times k$, with $\barbx_i(t)=\barby_i(t)-\bfeta_i(t), i=1,\ldots,k$. 

\noindent This expression for $\bB_{n,0}$ can further be written as $\bB_{n,0}=\sum_{i_1=1}^k\sum_{i_2=1}^kh_{i_1i_2}\int_{\calT}\barbx_{i_1}(t)\barbx_{i_2}(t)^\T dt$. It is then straightforward  to compute the expectation of $\bB_{n,0}$ as
$\bOmega_n=\E(\bB_{n,0})=
\int_{\calT}\sum_{i=1}^kh_{ii}[\bGamma_i(t,t)/n_i]dt$ $=\sum_{i=1}^kh_{ii}\bSigma_i/n_i$,
where $\bSigma_i = \int_{\calT}\bGamma_i(t,t)dt,i=1,\ldots,k$. Let $\hbSigma_i = \int_{\calT}\hbGamma_i(t,t)dt$.  It is evident that  the unbiased estimator of $\bOmega_n$ is $\hbOmega_n=\sum_{i=1}^kh_{ii}\hbSigma_i/n_i$, where the matrix of sample covariance functions $\hbGamma_i(t,t),i=1,\ldots,k$ is given in (\ref{kmeancov.sec2}). Let $\bE_n=\hbOmega_n$, then we can treat $\bE_n$ as the global variation matrix due to error. 

\begin{remark}
	When $\bC=(\bI_{k-1},-\bm{1}_{k-1})$ and $\bC_0(t)=\bm{0}$ for all $t\in\calT$,  the entries of $\bm{H}_n$ are given by $h_{ii}=n_{i}(n-n_i)/n$ for $i=1,\ldots,k$, and $h_{i_1i_2}=-n_{i_1} n_{i_2}/n$ for $i_1\neq i_2$ where $n=\sum_{i=1}^kn_i$. Consequently, $\bB_n=\int_{\calT}\sum_{i=1}^kn_i[\barby_i(t)-\barby(t)][\barby_i(t)-\barby(t)]^\T dt$, where $\barby(t)=n^{-1}\sum_{i=1}^k n_i\barby_i(t)$ represents  the vector of sample grand mean function. This formulation matches the matrix $\bm{H}$ defined in \cite{gorecki2017multivariate} and the matrix $\bB_n$ defined in \cite{Zhu2021}. However, the resulting $\bE_n=\int_{\calT}\sum_{i=1}^k[n(n_i-1)]^{-1}(n-n_i)\sum_{j=1}^{n_i}[\by_{ij}(t)-\barby_i(t)][\by_{ij}(t)-\barby_i(t)]^\T dt$ can be interpreted as an adjusted version of the within-group variation matrix defined in \cite{gorecki2017multivariate} and \cite{Zhu2021}. This adjustment is used to ensure that $\E(\bB_n)=\E(\bE_n)$ under $\H_0$.
\end{remark}
Since both $\bB_n$ and $\bE_n$ are symmetric and nonnegative-definite, following the ideas in \cite{harrar2012modified, xiao2016modified, zeng2024robust}, it is reasonable to approximate their distributions by two matrices with Wishart distributed. Let $W_p(m,\bV)$ denote a Wishart distribution of $m$ degrees of freedom and with covariance matrix $\bV:p\times p$. We expect to approximate the distributions of $\bB_n$ and $\bE_n$ by the distributions of the following Wishart random matrices: $\bB\sim W_p(d_B,\bOmega_n/d_B)$ and $\bE\sim W_p(d_E,\bOmega_n/d_E)$, where $d_B$ and $d_E$ are the approximate degrees of freedom for $\bB_n$ and $\bE_n$, respectively. The approximate degrees of freedom $d_B$ and $d_E$ can be determined via matching the total variations of $\bB_n$ and $\bB$ and those of $\bE_n$ and $\bE$. Note that the total variation of a random matrix $\bX=(x_{ij}):m\times m$ is the sum of the variances of all the entries of $\bX$, that is, $V(\bX)=\E\left[\tr(\bX-\E\bX)^2\right]=\sum_{i=1}^m\sum_{j=1}^m\Var(x_{ij})$, where and throughout $\tr(\bA)$ denotes the trace of a matrix $\bA$.  However, the resulting $d_B$ and $d_E$ are not affine-invariant so that the modified tests based on $\bB_n$ and $\bE_n$ will not be affine-invariant. To address this issue, if $\bB\sim W_p(d_B,\bOmega_n/d_B) \mbox{ and } \bE\sim W_p(d_E,\bOmega_n/d_E)$ hold, then we have $\tbB=\bOmega_n^{-1/2}\bB\bOmega_n^{-1/2}\sim W_p(d_B,\bI_p/d_B)$ and $\tbE=\bOmega_n^{-1/2}\bE\bOmega_n^{-1/2}\sim W_p(d_E,\bI_p/d_E)$. This implies that the approximate degrees of freedom $d_B$ and $d_E$ for $\bB_n$ and $\bE_n$ are also the approximate degrees of freedom of $\tbB_n=\bOmega_n^{-1/2}\bB_n\bOmega_n^{-1/2}$ and $\tbE_n=\bOmega_n^{-1/2}\bE_n\bOmega_n^{-1/2}$. Therefore, $d_B$ and $d_E$ can be determined via matching the total variations of $\tbB_n$ and $\tbB$, and those of $\tbE_n$ and $\tbE$, respectively.
That is, we need to solve the following two equations for $d_B$ and $d_E$:
\be\label{matequ.sec2}
V(\tbB)=V(\tbB_n),\; \mbox{ and }\; V(\tbE)=V(\tbE_n).
\ee

Let $\gamma_{i,h\ell}(s,t)$ be the $(h,\ell)$-th entry of $\bGamma_i(s,t),s,t\in\calT$. For simplicity of notation,  throughout this paper, for any matrix of covariance functions $\bGamma_i(s,t)$, $s,t\in\calT,i=1,\ldots,k$, we write $\tr(\bGamma_i)=\int_{\calT}\tr\left[\bGamma_i(t,t)\right]dt=\int_{\calT}\sum_{h=1}^p\gamma_{i,hh}(t,t)dt$;  
for any two covariance function matrices $\bGamma_{i_1}(s,t)$ and $\bGamma_{i_2}(s,t), s,t\in\calT, i_1,i_2=1,\ldots,k$, we write $I_{i_1i_2}=\int_{\calT^2}\tr[\bGamma_{i_1}(s,t)]\tr[\bGamma_{i_2}(s,t)]dsdt=\int_{\calT^2}\sum_{h=1}^p\sum_{\ell=1}^p\allowbreak\gamma_{i_1,hh}(s,t)\gamma_{i_2,\ell\ell}(s,t)dsdt$, and  $T_{i_1i_2}=\tr(\bGamma_{i_1}\bGamma_{i_2})=\int_{\calT^2}\tr\left[\bGamma_{i_1}(s,t)\bGamma_{i_2}(s,t)\right]dsdt=\int_{\calT^2}\sum_{h=1}^p\sum_{\ell=1}^p\allowbreak\gamma_{i_1,h\ell}(s,t)\gamma_{i_2,h\ell}(s,t)dsdt$. The following theorem provides the solution of (\ref{matequ.sec2}) and its proof is presented in Appendix.

\begin{theorem}\label{dBdEsol.thm}
	For each $i=1,\ldots,k$,  we assume that the vector of subject-effect functions $\bx_{ij}(t)=\by_{ij}(t)-\bm{\eta}_i(t),j=1,\ldots,n_i$ are identically and  independently distributed (i.i.d.).  Let $\bx^*_{ij}(t)=\bOmega_n^{-1/2}\bx_{ij}(t)$, $j=1,\ldots,n_{i};\;i=1,\ldots,k$, and $\bGamma^*_i(s,t)=\bOmega_n^{-1/2}\bGamma_{i}(s,t)\bOmega_n^{-1/2},i=1,\ldots,k$, then the solution of (\ref{matequ.sec2}) is given by
	\be\label{dB.sec2}
	d_B=\frac{p(p+1)}{\sum_{i=1}^kh_{ii}^2\calK_4(\bx^*_{i 1})/n_i^3+\sum_{i_1=1}^k\sum_{i_2=1}^kh_{i_1i_2}^2(I^*_{i_1i_2}+T_{i_1i_2}^*)/n_{i_1} n_{i_2}},
	\ee
	and 
	\be\label{dE.sec2}
	d_E=\frac{p(p+1)}{\sum_{i=1}^kh_{ii}^2\calK_4(\bx^*_{i1})/n_i^3+\sum_{i=1}^kh_{ii}^2(I^*_{ii}+T^*_{ii})/[n_{i}^2(n_i-1)]},
	\ee
	where  $\calK_4(\bx^*_{i 1})=\int_{\calT^2}\E[\bx^*_{i1}(s)^\T\bx^*_{i1}(t)\bx^*_{i1}(s)^\T\bx^*_{i1}(t)]dsdt-\tr(\bSigma^{*2}_i)-I_{ii}^*-T_{ii}^*$, with $\bSigma_i^*=\int_{\calT}\bGamma^*_i(t,t)dt$, $I^*_{i_1i_2}=\int_{\calT^2}\tr[\bGamma_{i_1}^*(s,t)]\tr[\bGamma_{i_2}^*(s,t)]dsdt$, and $T^*_{i_1i_2}=\tr(\bGamma_{i_1}^*\bGamma_{i_2}^*),i_1,i_2=1,\ldots,k$.
\end{theorem}

Now we have $d_B\bB_n\sim W_p(d_B,\bOmega_n)$ and $d_E\bE_n\sim W_p(d_E,\bOmega_n)$, approximately. Let $\bM_1=d_B\bB_n$ and $\bM_2=d_E\bE_n$. Then we can propose the following three test statistics for MFD, namely the modified functional Wilks (MFW) test statistic, the modified functional Lawley--Hotelling (MFLH) test statistic, and the modified functional Pillai (MFP) test statistic:
\begin{equation}\label{teststa.equ}
		T_{\MFW}=\frac{\det(\bM_2)}{\det(\bM_1+\bM_2)}, \;T_{\MFLH}=\tr(\bM_1\bM_2^{-1}),\;\mbox{ and }	T_{\MFP}=\tr[\bM_1(\bM_1+\bM_2)^{-1}],
\end{equation}
where $\det(\bA)$ denotes  the determinant of a square matrix $\bA$.


\begin{remark}\label{uni.rem}
	When $p=1$, the three test statistics coincide. Moreover, $\bB_n(t)$ in (\ref{Bn.sec2}) matches the value of $\mbox{SSH}_n(t)$ in \cite{smaga2019linear}, while $\bE_n$ resembles the denominator of $F_n$ in Equation (3) of   \cite{smaga2019linear}, though it is not identical. Consequently, the three proposed test statistics are comparable to the $F$-type test introduced by \cite{zhang2013analysis}, whose null distribution can be approximated by an 
	$F$-distribution.
\end{remark}
\begin{remark}
	Unlike the multivariate data scenario, even when the $k$ functional samples follow Gaussian processes, under the homogeneity assumption and the null hypothesis in (\ref{GLHTH0.sec2}),   the exact distributions of $\bB_n$ and $\bE_n$ cannot be determined.  However, at each time point $t\in\calT$, setting $\bB_n(t)=\bX(t)^\T\bm{H}_n\bX(t)$ and $\bE_n(t)=\sum_{i=1}^kh_{ii}\hbGamma(t,t)/n_i$, it can be shown that, under the homogeneity assumption and the null hypothesis in (\ref{GLHTH0.sec2}),  the pointwise variation matrices $\bB_n(t)$ and $\bE_n(t)$ follow Wishart distributions. 
	Nevertheless, simulation results indicate that our modified tests continue to perform well even when the functional samples are non-Gaussian.
\end{remark}

\subsection{Properties of the Proposed Tests}
The proposed tests possess several desirable invariance properties. This subsection introduces two key invariance properties of the tests. First, the proposed tests  $T_{\MFW}, T_{\MFLH}$, and $T_{\MFP}$ (\ref{teststa.equ}) are  affine invariant, meaning that the results of the statistical inference remain unaffected by affine transformations of the data.  This property is particularly useful in practice since the functional observations $\by_{ij}(t)$ are often recentered or rescaled before conducting an inference. These transformations are special cases of the affine transformation defined in (\ref{afftrans.equ}).
\begin{proposition}\label{afftrans1.prop}
	The proposed tests, $T_{\MFW}, T_{\MFLH}$, and $T_{\MFP}$ (\ref{teststa.equ}) are invariant under the following affine-transformation: 
	\be\label{afftrans.equ}
	\by^0_{ij}(t) = \bA \by_{ij}(t) + \bb(t), j=1,\ldots,n_i; i=1,\ldots,k,
	\ee
	where $\bA$ is any nonsingular matrix and $\bb(t)$ is any given function.
\end{proposition}
Besides, the proposed tests  $T_{\MFW}, T_{\MFLH}$, and $T_{\MFP}$ possess another invariance property. Note that for the hypotheses in  (\ref{GLHTH0.sec2}),  it is evident that the contrast matrix $\bC$ is not unique for the same hypothesis test. For instance,  in Example~\ref{fmanova.ex}, one valid choice for the one-way FMANOVA problem is $\bC=(\bI_{k-1},-\bm{1}_{k-1})$, while another equally valid choice is $\bC=(-\bm{1}_{k-1}, \bI_{k-1})$. Therefore, it is important to construct a test that remains invariant under non-singular transformations of the coefficient matrix $\bC$ and the constant matrix $\bC_0(t)$. 
\begin{proposition}\label{afftrans2.prop}
	The proposed tests, $T_{\MFW}, T_{\MFLH}$, and $T_{\MFP}$ (\ref{teststa.equ}) are invariant  when the coefficient matrix  $\bC$ and the constant matrix $\bC_0(t)$ are transformed as follows: 
	\be\label{lineartrans.equ}
	\bC: \bC\to \bP\bC, \;\mbox{ and }\;\bC_0(t):\bC_0(t)\to\bP\bC_0(t),
	\ee
	where $\bP$ is any nonsingular matrix of size $q\times q$.
\end{proposition}
The proofs of Propositions~\ref{afftrans1.prop} and \ref{afftrans2.prop} are provided in Appendix.

\subsection{Implementation}
The exact distributions of the three proposed test statistics (\ref{teststa.equ}) take quite complicated forms. Fortunately, they can be approximated by the $F$-approximation (\citealt{rao1951asymptotic}),  the chi-squared asymptotic expansion (\citealt{harrar2012modified}) or the normal-based asymptotic expansion  (\citealt{fujikoshi1975asymptotic}). \cite{zeng2024robust} noted that the $F$-approximation provides a more accurate estimation compared to the other two approximation methods. Additionally, as mentioned in Remark~\ref{uni.rem}, the $F$-distribution is used to approximate the test statistics in \cite{zhang2013analysis}, which is comparable to our proposed test statistics in the special case where $p=1$. Therefore, it is reasonable to utilize the $F$-approximation in our approach. 


Denote $\nu_1=(|d_B-p|-1)/2$ and $\nu_2=(d_E-p-1)/2$. Set $s=\min(p,d_B)$.  Let $\alpha$ denote the significance level and $F_{d_1,d_2}(\alpha)$ denote the upper $100\alpha$ percentile of the $F$-distribution with $d_1$ and $d_2$ degrees of freedom. The three tests can be conducted using $F$-approximation as follows:
\begin{itemize}
	\item  The $F$-approximation of $T_{\MFW}$ is given by 
	\[
	F_{\MFW}=\frac{\theta_1\theta_2-\theta_3}{pd_B}\frac{1-T_{\MFW}^{1/\theta_1}}{T_{\MFW}^{1/\theta_1}},
	\]
	where $\theta_1^2=(p^2d_B^2-4)/(p^2+d_B^2-5)$ if $p^2+d_B^2-5>0$ and $\theta_1=1$ otherwise, $\theta_2=d_E-(p-d_B+1)/2$, and $\theta_3=pd_B/2-1$. The MFW test will reject the null hypothesis if $F_{\MFW}$ is greater than $F_{pd_B,\theta_1\theta_2-\theta_3}(\alpha)$.
	\item  When $\nu_2\leq 0$, the $F$-approximation of $T_{\MFLH}$ is 
	\[
	F_{\MFLH} = 2(s\nu_2+1)[s^2(2\nu_1+s+1)]^{-1}T_{\MFLH},
	\]
	and the MFLH test will reject the null hypothesis if $F_{\MFLH}$ is greater than $F_{s(2\nu_1+s+1),2(s\nu_2+1)}(\alpha)$.\\
	When $\nu_2>0$,  the $F$-approximation of $T_{\MFLH}$ is 
	\[
	F_{\MFLH} = \Big(4+\frac{pd_B+2}{\phi_2-1}\Big)\frac{T_{\MFLH}}{pd_B\phi_1},
	\]
  where $\phi_1=[2+(pd_B+2)/(\phi_2-1)]/(2\nu_2)$ and $\phi_2=(p+2\nu_2)(d_B+2\nu_2)/[2(2\nu_2+1)(\nu_2-1)]$. The MFLH test will reject the null hypothesis if $F_{\MFLH}$ is greater than $F_{pd_B,4+(pd_B+2)/(\phi_2-1)}(\alpha)$.
  \item The $F$-approximation of  $T_{\MFP}$  is 
  \[
  F_{\MFP}=\frac{2\nu_2+s+1}{2\nu_1+s+1}\frac{T_{\MFP}}{s-T_{\MFP}},
  \]
  and the MFP test will reject the null hypothesis if $F_{\MFP}$ is greater than $F_{s(2\nu_1+s+1),s(2\nu_2+s+1)}(\alpha)$.
\end{itemize}

To implement the three proposed tests, we shall estimate the approximate degrees of freedom $d_B$ and $d_E$  as provided in Theorem~\ref{dBdEsol.thm}. Specifically, this involves estimating $\calK_4(\bx^*_{i 1}), \tr(\bSigma_i^{*2}),i=1,\ldots,k$, $I_{i_1i_2}^*$ and $T_{i_1i_2}^*,i_1,i_2=1,\ldots,k$ properly. These terms can be directly approximated by replacing $\bGamma_i^*(s,t),i=1,\ldots,k$ with its natural estimator, $\hbGamma^*_i(s,t)=\hbOmega_n^{-1/2}\hbGamma_{i}(s,t)\hbOmega_n^{-1/2}$, $i=1,\ldots,k$.  However, it is well known that $\tr(\hbGamma^2)$ and $\tr^2(\hbGamma)$ are poor estimators of $\tr(\bGamma^2)$ and $\tr^2(\bGamma)$, respectively, where $\hbGamma(s,t),s,t\in\calT$ denotes the usual matrix of sample covariance functions, which serves as an unbiased estimator of the matrix of covariance functions $\bGamma(s,t),s,t\in\calT$.  Furthermore, based on simulation results in \cite{Zhu2021,zhu2023ksampFMD} and additional pre-simulation studies, bias introduced by these natural estimators becomes significant, particularly when the MFD exhibit weak correlations. To address this issue, in this study, we adopt the U-statistics-based estimation approach utilized in \cite{li2012two}, which traces back to earlier works by \cite{glasser1961unbiased,glasser1962estimators}. Throughout this paper, we use $\sum^*$ to denote summation over mutually distinct indices. For example, $\sum_{i_1,i_2,i_3}^*$ means summation over $\{(i_1,i_2,i_3):i_1\neq i_2,i_2\neq i_3, i_3\neq i_1\}$. 

Let $\tilde{\by}_{ij}(t)=\bOmega_n^{-1/2}\by_{ij}(t),j=1,\ldots,n_i;i=1,\ldots,k$. Consequently, we have $\Cov[\tilde{\by}_{i1}(s),\tilde{\by}_{i1}(t)]=\bOmega_n^{-1/2}\bGamma_i(s,t)\bOmega_n^{-1/2}=\bGamma_i^*(s,t)$. By applying Lemma~\ref{Ustat.lem}  in Appendix, the unbiased estimators for $I^*_{i_1i_2},T^*_{i_1i_2},\;i_1,i_2=1,\ldots,k$ and $\tr(\bSigma_i^{*2}),i=1,\ldots,k$ can be obtained by substituting $\tilde{\by}_{ij}(t)$. Unfortunately, $\bOmega_n$ is unknown but can be replaced with its unbiased estimator, $\hbOmega_n$. Therefore, we propose the following  bias-reduced estimators of $I^*_{ii}$, $T_{ii}^*$, and $\tr(\bSigma_i^{*2}),i=1,\ldots,k$ by substituting $\by^*_{ij}(t)=\hbOmega_n^{-1/2}\by_{ij}(t)$ for $\by_{ij}(t),j=1,\ldots,n_i;i=1,\ldots,k$ as follows:
\begin{equation}\label{Iii.equ}
		\begin{array}{rcl}
	\hI^*_{ii}&=&[n_i(n_i-1)]^{-1}\sum_{j_1\neq j_2}\int_{\calT^2}\delta_{i,j_1j_1}(t,s)\delta_{i,j_2j_2}(t,s)dsdt\\
		&-&2[n_i(n_i-1)(n_i-2)]^{-1}\sum_{j_1,j_2,j_3}^*\int_{\calT^2}\delta_{i,j_1j_1}(t,s)\delta_{i,j_2j_3}(t,s)dsdt\\
		&+&[n_i(n_i-1)(n_i-2)(n_i-3)]^{-1}\sum_{j_1,j_2,j_3,j_4}^*\int_{\calT^2}\delta_{i,j_1j_2}(t,s)\delta_{i,j_3j_4}(t,s)dsdt,\\
		\hT^*_{ii}	&=&[n_i(n_i-1)]^{-1}\sum_{j_1\neq j_2}\int_{\calT^2}\delta_{i,j_1j_2}(t,s)\delta_{i,j_2j_1}(t,s)dsdt\\
		&-&2[n_i(n_i-1)(n_i-2)]^{-1}\sum_{j_1,j_2,j_3}^*\int_{\calT^2}\delta_{i,j_1j_2}(t,s)\delta_{i,j_3j_1}(t,s)dsdt\\
		&+&[n_i(n_i-1)(n_i-2)(n_i-3)]^{-1}\sum_{j_1,j_2,j_3,j_4}^*\int_{\calT^2}\delta_{i,j_2j_3}(t,s)\delta_{i,j_4j_1}(t,s)dsdt,\\
		\widehat{\tr(\bSigma_i^{*2})}	&=&[n_i(n_i-1)]^{-1}\sum_{j_1\neq j_2}\int_{\calT^2}\delta_{i,j_1j_2}(s,t)\delta_{i,j_2j_1}(t,s)dsdt\\
		&-&2[n_i(n_i-1)(n_i-2)]^{-1}\sum_{j_1,j_2,j_3}^*\int_{\calT^2}\delta_{i,j_1j_2}(s,t)\delta_{i,j_3j_1}(t,s)dsdt\\
		&+&[n_i(n_i-1)(n_i-2)(n_i-3)]^{-1}\sum_{j_1,j_2,j_3,j_4}^*\int_{\calT^2}\delta_{i,j_2j_3}(s,t)\delta_{i,j_4j_1}(t,s)dsdt,
			\end{array}
\end{equation}
where $\delta_{i,j_1j_2}(s,t)=\by_{ij_1}(s)^\T\hbOmega_n^{-1}\by_{ij_2}(t),s,t\in\calT,j_1,j_2=1,\ldots,n_i.$

To estimate $\calK_4(\bx^*_{i1}), i=1,\ldots,k$ properly, we adopt the approach used by \cite{himeno2014estimations} and \cite{zhu2023ksampFMD}, and propose the following estimator: 
\begin{equation}\label{calK4.equ}
			\widehat{\calK_4(\bx^*_{i1})}=  (n_i-1)^{-1}\sum_{j=1}^{n_{i}}\int_{\calT^2}\{[\by_{ij}(s)-\barby_{i}(s)]^\T\hbOmega_n^{-1}[\by_{ij}(t)-\barby_i(t)]\}^2dsdt-\widehat{\tr(\bSigma_i^{*2})} -\hI_{ii}^*-\hT_{ii}^*,
\end{equation}
where $\widehat{\tr(\bSigma_i^{*2})}, \hI_{ii}^*$, and $\hT_{ii}^*,i=1,\ldots,k,$ are given in (\ref{Iii.equ}). Similarly, the bias-reduced estimators of $I^*_{i_1i_2}$ and $T^*_{i_1i_2}, i_1\neq i_2$ also can be obtained by applying Lemma~\ref{Ustat.lem} in Appendix, as shown below:
\begin{equation}\label{Iij.equ}
	\begin{array}{rcl}
		\hI^*_{i_1i_2} &=& \int_{\calT^2}\tr[\hbGamma^*_{i_1}(s,t)]\tr[\hbGamma^*_{i_2}(s,t)]dsdt=\int_{\calT^2}\tr[\hbOmega_n^{-1}\hbGamma_{i_1}(s,t)]\tr[\hbOmega_n^{-1}\hbGamma_{i_2}(s,t)]dsdt,\mbox{ and }\\
		\hT^*_{i_1i_2} &=& \tr(\hbGamma^*_{i_1}\hbGamma^*_{i_2})=\int_{\calT^2}\tr[\hbOmega_n^{-1}\hbGamma_{i_1}(s,t)\hbOmega_n^{-1}\hbGamma_{i_2}(s,t)]dsdt.
	\end{array}
\end{equation}
Therefore, the bias-reduced estimators of $d_B$ (\ref{dB.sec2}) and $d_E$ (\ref{dE.sec2}) are given by 
\be\label{hdB.sec2}
\hd_B=\frac{p(p+1)}{\sum_{i=1}^kh_{ii}^2[\widehat{\calK_4(\bx^*_{i1})}/n_i^3+(\hI_{ii}^*+\hT_{ii}^*)/n_i^2]+\sum_{i_1\neq i_2} h_{i_1i_2}^2(\hI^*_{i_1i_2}+\hT_{i_1i_2}^*)/(n_{i_1} n_{i_2})}, 
\ee
and
\be\label{hdE.sec2}
\hd_E=\frac{p(p+1)}{\sum_{i=1}^kh_{ii}^2\{\widehat{\calK_4(\bx^*_{i1})}/n_i^3+(\hI^*_{ii}+\hT^*_{ii})/[n_{i}^2(n_i-1)]\}},
\ee
respectively, where $\hI^*_{i_1i_2},\hT^*_{i_1i_2},i_1,i_2=1,\ldots,k$, and $\widehat{\calK_4(\bx^*_{i1})},i=1,\ldots,k$ are provided in  (\ref{Iii.equ}), (\ref{calK4.equ}), and (\ref{Iij.equ}).
Let $\hd_B$ and $\hd_E$ be the estimated degrees of freedom obtained from (\ref{hdB.sec2}) and (\ref{hdE.sec2}), respectively, and $\hbM_1=\hd_B\bB_n$ and $\hbM_2=\hd_E\bE_n$. Then the proposed tests can be conducted by employing the $F$-approximation introduced above.

\section{Simulation Studies}\label{Simu.sec}
In this section, we perform simulation studies to assess the finite-sample performance of the three proposed tests: $T_{\MFW}$, $T_{\MFLH}$, and $T_{\MFP}$. We consider a fixed number of samples with $k = 4$ and examine three cases for $\bn = [n_1, n_2, n_3, n_4]$: $\bn_1 = [10,10,10,10]$, $\bn_2 = [10,12,12,15]$, and $\bn_3 = [15,15,25,25]$. For computational efficiency, the generated functions are observed at $M = 80$ equidistant points in the closed interval $\calT = [0,1]$, as \cite{zhang2014one} suggests that $M = 50 \sim 1000$ is often sufficient for various functional data purposes.   Although larger values of $M$ could be used, they often increase computational cost without significantly improving test performance. The effect of $M$ is discussed in \cite[Sec.4.5.6]{zhang2013analysis}.

Following simulation studies of  \cite{smaga2019linear}, \cite{Qiu2021}, and \cite{munko2024multiple},  the $k$ multivariate functional samples are generated using the model: $\by_{ij}(t)=\bfeta_i(t)+\sum_{r=1}^q\sqrt{\lambda_{ir}}\varepsilon_{ijr}\bphi_r(t), j=1,\ldots,n_i;$ $i=1,\ldots,4$. Here, $\varepsilon_{ijr}$ are i.i.d. random variables, and the orthonormal basis functions $\bm{\phi}_r(t)$ along with the variance components $\lambda_{ir}$, in descending order, are used to define the covariance matrix functions $\bGamma_i(s,t) = \sum_{r=1}^q \lambda_{ir} \bm{\phi}_r(s) \bm{\phi}_r(t)^{\top}$ for $i=1, \ldots, 4$. The goal is to compare the finite-sample performance of $T_{\MFW}$, $T_{\MFLH}$ and $T_{\MFP}$ with other tests when the dimension $p$ of the generated MFD is reasonably large, and we set $p=6$.   The vector of mean functions for the first group, $\bfeta_1(t)=[\eta_{11}(t),\ldots,\eta_{16}(t)]^\top$, is defined as $\eta_{11}(t)=[\sin(2\pi t^2)]^5$, $\eta_{12}(t)=[\cos(2\pi t^2)]^5,\eta_{13}(t)=t^{1/3}(1-t)-5,\eta_{14}(t)=\sqrt{5}t^{2/3}\exp(-7t),\eta_{15}(t)=\sqrt{13t}\exp(-13t/2)$, and $\eta_{16}(t)=1+2.3t+3.4t^2+1.5t^3$.
To define  the mean functions for the other three groups, we set $\bfeta_2(t)=\bm{\eta}_1(t)$ and $\bfeta_3(t)=\bm{\eta}_4(t)$. Specifically,  $\bfeta_{3\ell}(t)=\bm{\eta}_{1\ell}(t),\ell=1,\ldots,5$ and $\eta_{36}(t)=(1+\delta/\sqrt{30})+(2.3+2\delta/\sqrt{30})t+(3.4+3\delta/\sqrt{30})t^2+(1.5+4\delta/\sqrt{30})t^3$, where $\delta$ controls the difference in the mean functions.

To define the matrices of  covariance functions $\bGamma_i(s,t)$, we set $\lambda_{ir}=\nu_{i}\rho^r,r=1,\ldots,q$, with $\rho=0.1,0.5,0.9$, representing high, moderate, and low correlations among the components of the simulated functional data. When $\rho$ is large, the eigenvalues $\lambda_{ir}$ decay more slowly, making the functional samples noisier.   
We consider two scenarios: S1: $\nu_1=\nu_2=\nu_3=\nu_4=1.5$; S2: $\nu_1=1.5,\nu_2=2,\nu_3=2.5,\nu_4=3$ for homoscedastic and heteroscedastic cases, respectively. 
For $\bm{\phi}_r(t)=[c_1\psi_r(t),\ldots,c_p\psi_r(t)]^\top$,   the basis functions are taken as $\psi_1(t)=1$, $\psi_{2r}(t)=\sqrt{2}\sin(2\pi rt)$, $\psi_{2r+1}(t)=\sqrt{2}\cos(2\pi rt),t\in\calT, r=1,\ldots,(q-1)/2$,  
and we let $c_{\ell}=\ell/(1^2+\cdots+p^2)^{1/2},\ell=1,\ldots,p$, so that $\sum_{\ell=1}^pc_{\ell}^2=1$ and $\int_{\calT}\bm{\phi}_r(t)^\top\bm{\phi}_r(t)dt=1$ holds for $r=1,\ldots,q$. We take $q=7$.  To generate Gaussian functional data we set $\varepsilon_{ijr}\iidsim N(0,1)$ (Model 1),  and we specify $\varepsilon_{ijr}\iidsim t_8/\sqrt{4/3}$ (Model 2) and $\varepsilon_{ijr}\iidsim (\chi_4^2-4)/(2\sqrt{2})$ (Model 3) to generate non-Gaussian functional data. 

Throughout this section, we set the nominal size $\alpha$ to $5\%$ and  conduct a total of  $N=1000$ simulation runs under each configuration.  To evaluate a test's overall ability to maintain the nominal size, we employ the average relative error (ARE) metric, as proposed by \cite{zhang2011tech}. The ARE value for a test is computed as follows:
$\ARE=100J^{-1}\sum_{j=1}^J|\halpha_j-\alpha|/\alpha$, where $\halpha_j,j=1,\ldots,J$ denote the empirical sizes under $J$ simulation settings. A lower ARE value reflects superior performance of the test in controlling size.

\subsection{Simulation 1: One-way FMANOVA}
We first focus on Example~\ref{BFH0.sec2} in Section~\ref{model.sec}, which involves testing homogeneity of several mean functions. The three proposed tests were compared to several competing tests introduced by \cite{gorecki2017multivariate}, \cite{Zhu2021}, \cite{zhu2023ksampFMD}, \cite{qiu2024tests}, and \cite{zhu2023general}. Specifically, the permutation tests proposed by \cite{gorecki2017multivariate}, which are based on a basis function representation of functional data,  use the Wilks', Lowley--Hotelling's, Pillay's and Roy's test statistics are denoted as W, LH, P, and R, respectively. The functional Lowley--Hotelling trace (FLH) tests proposed by \cite{Zhu2021}, with the naive and bias-reduced methods are denoted as $T_{\FLH}^\N$ and $T_{\FLH}^\B$, respectively.  The global tests proposed by \cite{zhu2023ksampFMD} for the heteroscedastic one-way MANOVA problem with MFD,  using  the naive and bias-reduced methods are denoted as $T_{\ZZC}^{\N}$ and $T_{\ZZC}^{\B}$, respectively. The two novel global testing statistics, derived by integrating or maximizing the pointwise Lawley--Hotelling trace test statistic proposed by \cite{qiu2024tests}, are denoted as $T_{\QFZ}$ and $T_{\QFZ}^{\max}$, respectively. Additionally, the test proposed by \cite{zhu2023general} for the GLHT problem under heteroscedasticity is referred to as $T_{\Z}$. We can employ $\bC=(\bI_3, -\bm{1}_3)$ and  $\bC_0(t)=\bm{0}$ for all $t\in\calT$ in (\ref{GLHTH0.sec2}) to implement the three new tests.

\begin{table}[!h]
	\centering
	\caption{Empirical sizes (in $\%$)  in Simulation 1 under homoscedastic case (S1).}
	\vspace{1mm}
	\setlength{\tabcolsep}{1pt} 
	\scalebox{0.9}{
		\begin{tabular*}{\textwidth}{@{\extracolsep{\fill}}cccccccccccccccc@{\extracolsep{\fill}}}
			\toprule
			\multicolumn{3}{l}{$\rho=0.1$} \\
			\midrule
			Model	 & $\bn$     & W     & LH    & P     & R     & $T_{\FLH}^\N$  & $T_{\FLH}^\B$  & $T_{\ZZC}^\N$  & $T_{\ZZC}^\B$ &$T_{\QFZ}$ &$T_{\QFZ}^{\max}$ & $T_{\Z}$ & $T_{\MFW}$  & $T_{\MFLH}$  & $T_{\MFP}$ \\
			\midrule
			\multirow{3}[0]{*}{1} & $\bn_1$    & 5.6   & 5.3   & 5.5   & 5.6   & 19.1  & 23.1  & 13.5  & 22.8  & 23.6  & 5.7   & 2.9   & 7.2   & 7.8   & 6.0 \\
			& $\bn_2$    & 4.8   & 5.0   & 5.2   & 4.3   & 14.1  & 16.7  & 9.7   & 16.0  & 16.7  & 5.6   & 3.4   & 5.4   & 5.9   & 5.2 \\
			& $\bn_3$    & 5.8   & 5.6   & 5.9   & 4.1   & 9.8   & 11.1  & 7.2   & 11.3  & 11.6  & 5.7   & 5.6   & 6.4   & 6.4   & 6.0 \\
			\multirow{3}[0]{*}{2} & $\bn_1$    & 5.6   & 5.3   & 6.0   & 4.4   & 18.2  & 23.7  & 11.4  & 23.8  & 24.4  & 3.9   & 2.2   & 6.3   & 6.5   & 5.1 \\
			& $\bn_2$    & 5.1   & 5.0   & 5.5   & 5.1   & 15.7  & 18.8  & 9.8   & 18.7  & 18.7  & 5.6   & 3.0   & 6.2   & 6.6   & 5.0 \\
			& $\bn_3$    & 4.5   & 4.3   & 4.9   & 4.5   & 9.0   & 10.7  & 6.3   & 10.3  & 10.4  & 5.4   & 3.6   & 4.8   & 4.7   & 4.3 \\
			\multirow{3}[0]{*}{3} & $\bn_1$    & 5.1   & 5.2   & 4.9   & 4.8   & 17.6  & 23.3  & 11.5  & 24.1  & 24.2  & 5.3   & 1.9   & 5.3   & 5.5   & 4.2 \\
			& $\bn_2$    & 4.6   & 4.4   & 4.7   & 4.8   & 13.0  & 15.7  & 7.9   & 15.0  & 15.7  & 4.4   & 2.5   & 4.6   & 4.6   & 3.3 \\
			& $\bn_3$    & 5.0   & 5.2   & 4.9   & 5.1   & 9.0   & 11.4  & 6.0   & 10.9  & 11.2  & 5.2   & 3.8   & 4.5   & 4.5   & 3.9 \\
			\midrule
			\multicolumn{2}{c}{ARE} & 7.33  & 6.44  & 8.22  & 8.67  & 178.89 & 243.33 & 85.11 & 239.78 & 247.78 & 11.56 & 38.44 & 17.56 & 22.00 & 14.67 \\
			
			\midrule
			\multicolumn{3}{l}{$\rho=0.5$} \\
			\midrule
			Model	 & $\bn$     & W     & LH    & P     & R     & $T_{\FLH}^\N$  & $T_{\FLH}^\B$  & $T_{\ZZC}^\N$  & $T_{\ZZC}^\B$ &$T_{\QFZ}$ &$T_{\QFZ}^{\max}$ & $T_{\Z}$ & $T_{\MFW}$  & $T_{\MFLH}$  & $T_{\MFP}$ \\
			\midrule
			\multirow{3}[0]{*}{1} & $\bn_1$    & 5.4   & 5.6   & 5.4   & 5.7   & 8.6   & 15.8  & 11.4  & 34.9  & 32.9  & 6.1   & 2.2   & 6.9   & 7.3   & 5.4 \\
			& $\bn_2$    & 5.5   & 5.1   & 5.6   & 4.0   & 7.3   & 11.5  & 7.8   & 23.8  & 21.6  & 4.3   & 2.9   & 6.2   & 6.3   & 5.8 \\
			& $\bn_3$    & 5.5   & 5.5   & 5.5   & 5.0   & 6.4   & 8.3   & 6.1   & 15.7  & 14.6  & 5.1   & 4.4   & 6.0   & 5.9   & 5.7 \\
			\multirow{3}[0]{*}{2} & $\bn_1$    & 5.6   & 5.8   & 5.5   & 4.4   & 7.8   & 15.1  & 9.1   & 35.7  & 32.6  & 4.9   & 1.3   & 5.8   & 6.0   & 5.2 \\
			& $\bn_2$    & 6.5   & 5.9   & 6.2   & 5.3   & 8.3   & 13.0  & 8.1   & 23.3  & 21.9  & 6.1   & 3.0   & 6.1   & 6.6   & 5.7 \\
			& $\bn_3$    & 4.4   & 4.4   & 4.3   & 4.7   & 5.4   & 8.7   & 5.1   & 12.7  & 12.0    & 5.0   & 4.0   & 4.7   & 5.2   & 4.4 \\
			\multirow{3}[0]{*}{3} & $\bn_1$    & 4.4   & 4.2   & 4.3   & 4.7   & 6.9   & 15.4  & 8.9   & 34.7  & 31.9  & 4.6   & 1.3   & 4.6   & 5.2   & 3.9 \\
			& $\bn_2$    & 4.7   & 4.6   & 4.9   & 4.4   & 6.4   & 10.8  & 5.4   & 21.9  & 19.4  & 3.5   & 2.0   & 4.9   & 5.1   & 4.0 \\
			& $\bn_3$    & 4.7   & 4.9   & 4.8   & 5.0   & 5.7   & 9.1   & 5.1   & 14.1  & 12.6  & 6.0   & 3.7   & 4.6   & 4.7   & 4.2 \\
			\midrule
			\multicolumn{2}{c}{ARE} & 11.78 & 10.67 & 10.89 & 8.44  & 39.56 & 139.33 & 48.89 & 381.78 & 343.33 & 13.33 & 44.89 & 16.00 & 17.56 & 14.00 \\
			\midrule
			\multicolumn{3}{l}{$\rho=0.9$}\\
			\midrule
			Model  & $\bn$     & W     & LH    & P     & R     & $T_{\FLH}^\N$  & $T_{\FLH}^\B$  & $T_{\ZZC}^\N$  & $T_{\ZZC}^\B$ &$T_{\QFZ}$ &$T_{\QFZ}^{\max}$ & $T_{\Z}$ & $T_{\MFW}$  & $T_{\MFLH}$  & $T_{\MFP}$ \\
			\midrule
			\multirow{3}[0]{*}{1} & $\bn_1$    & 5.8   & 5.7   & 5.9   & 5.6   & 2.5   & 10.7  & 6.6   & 51.9  & 40.2  & 5.6   & 0.2   & 6.8   & 6.6   & 6.3 \\
			& $\bn_2$    & 5.1   & 5.2   & 5.1   & 4.6   & 2.0   & 8.4   & 4.3   & 35.2  & 25.6  & 5.7   & 1.3   & 5.6   & 5.3   & 5.5 \\
			& $\bn_3$    & 5.0   & 5.1   & 4.9   & 4.9   & 3.1   & 7.4   & 3.0   & 17.4  & 14.7  & 4.7   & 1.8   & 5.3   & 5.7   & 5.1 \\
			\multirow{3}[0]{*}{2}  &$\bn_1$    & 5.6   & 5.6   & 5.6   & 5.3   & 2.2   & 10.8  & 5.6   & 50.9  & 38.9  & 5.3   & 0.4   & 6.4   & 6.1   & 5.8 \\
			& $\bn_2$    & 5.6   & 5.7   & 5.8   & 5.9   & 3.1   & 9.4   & 3.9   & 32.8  & 24.3  & 6.6   & 1.4   & 5.9   & 5.7   & 5.8 \\
			& $\bn_3$    & 4.6   & 4.7   & 4.6   & 5.6   & 3.0   & 6.6   & 2.5   & 17.0  & 13.5  & 6.7   & 2.1   & 4.7   & 4.8   & 4.7 \\
			\multirow{3}[0]{*}{3} & $\bn_1$    & 4.6   & 4.6   & 4.3   & 4.6   & 1.3   & 8.6   & 4.0   & 48.7  & 36.9  & 5.2   & 0.2   & 3.9   & 4.2   & 3.6 \\
			& $\bn_2$    & 4.7   & 4.8   & 4.9   & 4.5   & 2.8   & 9.3   & 3.6   & 31.4  & 22.9  & 4.6   & 1.2   & 4.5   & 4.2   & 4.1 \\
			& $\bn_3$    & 5.4   & 5.4   & 5.3   & 5.7   & 3.0   & 7.4   & 2.9   & 17.5  & 12.6  & 6.8   & 2.1   & 5.2   & 5.1   & 4.9 \\
			\midrule
			\multicolumn{2}{c}{ARE} & 8.00  & 8.00  & 8.89  & 10.00 & 48.89 & 74.67 & 28.89 & 572.89 & 410.22 & 16.89 & 76.22 & 15.78 & 14.00 & 13.78 \\
			\bottomrule
		\end{tabular*}
		\label{size.tab1}}
\end{table}%

Table~\ref{size.tab1} shows the empirical sizes (in \%) of all the considered tests under the one-way FMANOVA problem (S1). The last row of each sub-table summarizes the ARE values corresponding to the three different $\rho$ values. Several key conclusions can be drawn from the results. First, the proposed new tests, $T_{\MFW}$, $T_{\MFLH}$ and $T_{\MFP}$, consistently demonstrate strong performance in terms of ARE values, regardless of the correlations among the components of the simulated functional data.  Specifically, when $\rho=0.1$, the ARE values are 17.56, 22.00, and 14.67 for $T_{\MFW}$, $T_{\MFLH}$ and $T_{\MFP}$, respectively; for $\rho=0.5$, the corresponding ARE values are 16.00, 17.56, and 14.00 for $T_{\MFW}$, $T_{\MFLH}$ and $T_{\MFP}$, respectively. When $\rho=0.9$, the ARE values are 15.78 for $T_{\MFW}$, 14.00 for $T_{\MFLH}$ and 13.78 for $T_{\MFP}$. Among these three proposed tests, in terms of overall ability to maintain the nominal size (ARE values), $T_{\MFP}$ performs slightly better than  $T_{\MFW}$ and  $T_{\MFLH}$  when the functional data are highly correlated ($\rho=0.1$), but it is slightly conservative under Model 3. The performances of the three tests are comparable under moderate ($\rho=0.5$) and low correlations ($\rho=0.9$). Both $T_{\MFW}$ and $T_{\MFLH}$ perform well for non-Gaussian data (Model 2 and Model 3), especially when the sample sizes are not too small ($\bn=\bn_2$ and $\bn_3$).  Second,  the four permutation-based tests proposed by \cite{gorecki2017multivariate} (W, LH, P, and R) perform consistently well, regardless of whether the functional data exhibit high correlation ($\rho=0.1$), moderate correlation ($\rho=0.5$), or low correlation ($\rho=0.9$).  Their performance remains robust regardless of whether the generated data are Gaussian or non-Gaussian. Third,  the FLH tests  proposed by \cite{Zhu2021} ($T_{\FLH}^\N$ and $T_{\FLH}^\B$) and the global tests from \cite{zhu2023ksampFMD} ($T_{\ZZC}^\N$ and $T_{\ZZC}^\B$) do not perform as well as \cite{gorecki2017multivariate}'s tests. Specifically, the tests with the naive method ($T_{\FLH}^\N$ and $T_{\ZZC}^\N$) become increasingly conservative as correlation decreases, while the tests with the bias-reduced method ($T_{\FLH}^\B$ and $T_{\ZZC}^\B$) tend to be overly liberal, especially when the data are highly or moderately correlated.  These findings align with the simulation results from \cite{Zhu2021} and \cite{zhu2023ksampFMD}, but the smaller sample sizes in this study lead to even poorer performance for these tests. Fourth, $T_{\QFZ}$ performs similarly to $T_{\ZZC}^\B$ as their test statistics are equivalent and are implemented in a similar manner. $T_{\QCZ}^{\max}$ generally performs well, particularly when $\rho=0.1$, outpacing the three new tests. Finally, the test developed by \cite{zhu2023general} is highly conservative in this simulation. This is attributed to the indiscriminate use of their adjustment coefficient $c_n$. The large value of $c_n$, which deviates significantly from 1, causes the test to behave conservatively.
\begin{table}[!h]
	\centering
	\caption{Empirical powers (in $\%$)  in Simulation 1 under homoscedastic case (S1) when $\bn=\bn_3$.}
	\vspace{1mm}
\setlength{\tabcolsep}{1pt} 
\scalebox{0.9}{
	\begin{tabular*}{\textwidth}{@{\extracolsep{\fill}}cccccccccccccccc@{\extracolsep{\fill}}}
		\toprule
		\multicolumn{3}{l}{$\rho=0.1$} \\
		\midrule
	Model  & $\delta$     & W     & LH    & P     & R     & $T_{\FLH}^\N$  & $T_{\FLH}^\B$  & $T_{\ZZC}^\N$  & $T_{\ZZC}^\B$ &$T_{\QFZ}$ &$T_{\QFZ}^{\max}$ & $T_{\Z}$ & $T_{\MFW}$  & $T_{\MFLH}$  & $T_{\MFP}$ \\
		\midrule
		\multirow{4}[0]{*}{1} & 0.1   & 10.5  & 10.1  & 10.8  & 7.9   & 18.5  & 20.8  & 14.7  & 20.1  & 20.0  & 27.4  & 8.7   & 11.9  & 11.6  & 10.8 \\
		& 0.2   & 38.0  & 38.9  & 37.3  & 34.4  & 51.4  & 53.6  & 45.8  & 53.3  & 53.0  & 96.7  & 35.7  & 38.9  & 39.9  & 36.5 \\
		& 0.3   & 82.5  & 83.4  & 80.9  & 83.3  & 90.5  & 91.4  & 87.8  & 91.8  & 91.7  & 100.0 & 80.3  & 83.1  & 84.3  & 80.2 \\
		& 0.4   & 98.9  & 99.2  & 98.7  & 99.6  & 99.7  & 99.7  & 99.6  & 99.8  & 99.7  & 100.0 & 98.7  & 99.0  & 99.3  & 98.2 \\
		\multirow{4}[0]{*}{2} & 0.1   & 9.8   & 9.9   & 10.2  & 8.1   & 16.9  & 19.0  & 12.3  & 18.0  & 18.3  & 25.9  & 7.9   & 10.3  & 10.3  & 9.8 \\
		& 0.2   & 36.8  & 36.6  & 36.0  & 35.6  & 49.8  & 53.1  & 41.5  & 52.9  & 52.4  & 95.3  & 31.9  & 35.7  & 36.6  & 33.5 \\
		& 0.3   & 81.0  & 81.6  & 79.8  & 81.2  & 89.0  & 90.8  & 85.0  & 90.7  & 90.7  & 100.0 & 77.6  & 79.8  & 80.7  & 76.8 \\
		& 0.4   & 98.6  & 98.7  & 98.3  & 99.2  & 99.6  & 99.6  & 99.1  & 99.6  & 99.6  & 100.0 & 97.8  & 98.2  & 98.3  & 97.8 \\
		\multirow{4}[0]{*}{3} & 0.1   & 9.6   & 9.5   & 9.2   & 8.1   & 17.4  & 20.0  & 12.3  & 19.8  & 19.6  & 29.0  & 7.5   & 9.3   & 9.6   & 8.1 \\
		& 0.2   & 38.6  & 38.7  & 39.5  & 36.5  & 51.3  & 55.0  & 43.7  & 54.6  & 54.0  & 94.7  & 35.0  & 37.1  & 37.2  & 35.1 \\
		& 0.3   & 82.1  & 82.7  & 80.9  & 82.7  & 89.7  & 90.8  & 85.0  & 91.0  & 91.0  & 100.0 & 76.6  & 78.5  & 80.2  & 74.1 \\
		& 0.4   & 98.5  & 98.8  & 98.1  & 99.3  & 99.3  & 99.4  & 99.0  & 99.3  & 99.5  & 100.0 & 97.3  & 97.4  & 97.9  & 95.2 \\
		\midrule
		\multicolumn{3}{l}{$\rho=0.5$} \\
		\midrule
			Model  & $\delta$     & W     & LH    & P     & R     & $T_{\FLH}^\N$  & $T_{\FLH}^\B$  & $T_{\ZZC}^\N$  & $T_{\ZZC}^\B$ &$T_{\QFZ}$ &$T_{\QFZ}^{\max}$ & $T_{\Z}$ & $T_{\MFW}$  & $T_{\MFLH}$  & $T_{\MFP}$ \\
		\midrule
		\multirow{4}[0]{*}{1} & 0.2   & 9.4   & 9.1   & 9.5   & 8.4   & 11.0  & 15.6  & 10.8  & 20.7  & 19.6  & 10.2  & 8.3   & 10.2  & 10.6  & 9.4 \\
		& 0.4   & 25.7  & 26.5  & 25.2  & 27.9  & 30.2  & 36.8  & 27.4  & 46.8  & 45.3  & 49.5  & 22.6  & 26.7  & 27.4  & 25.2 \\
		& 0.6   & 65.4  & 66.0  & 63.1  & 70.1  & 70.6  & 76.1  & 68.3  & 82.2  & 81.8  & 94.3  & 61.3  & 66.6  & 67.9  & 63.3 \\
		& 0.8   & 92.0  & 93.3  & 91.0  & 96.2  & 94.6  & 96.2  & 94.3  & 97.9  & 97.6  & 100.0 & 91.7  & 93.1  & 93.9  & 91.1 \\
		\multirow{4}[0]{*}{2} & 0.2   & 8.4   & 8.4   & 8.3   & 7.5   & 10.3  & 13.9  & 7.9   & 18.7  & 17.4  & 10.3  & 6.0   & 8.3   & 8.8   & 7.6 \\
		& 0.4   & 26.2  & 26.7  & 26.3  & 24.7  & 30.4  & 36.3  & 27.3  & 43.7  & 42.5  & 51.5  & 21.1  & 27.0  & 27.7  & 25.3 \\
		& 0.6   & 62.8  & 64.0  & 61.1  & 68.5  & 66.8  & 73.7  & 65.2  & 81.8  & 81.0  & 94.9  & 58.2  & 62.0  & 64.2  & 59.5 \\
		& 0.8   & 92.7  & 93.7  & 91.5  & 95.4  & 94.2  & 96.0  & 93.7  & 97.4  & 97.2  & 100.0 & 91.6  & 91.7  & 93.3  & 89.8 \\
		\multirow{4}[0]{*}{3} & 0.2   & 9.1   & 9.1   & 9.2   & 7.2   & 10.3  & 14.3  & 8.8   & 21.3  & 19.5  & 10.7  & 6.5   & 8.6   & 8.7   & 8.1 \\
		& 0.4   & 27.5  & 28.2  & 26.9  & 26.1  & 31.1  & 37.7  & 29.0  & 47.4  & 45.1  & 51.9  & 22.7  & 27.0  & 28.2  & 24.7 \\
		& 0.6   & 64.2  & 65.6  & 62.2  & 71.0  & 69.4  & 74.8  & 66.1  & 82.7  & 81.5  & 93.9  & 59.7  & 62.2  & 65.6  & 58.9 \\
		& 0.8   & 93.1  & 93.4  & 91.7  & 96.5  & 94.3  & 95.9  & 93.0  & 97.7  & 97.3  & 100.0 & 90.2  & 91.2  & 92.5  & 89.4 \\
		\midrule
		\multicolumn{3}{l}{$\rho=0.9$}\\
		\midrule
			Model   & $\delta$       & W     & LH    & P     & R     & $T_{\FLH}^\N$  & $T_{\FLH}^\B$  & $T_{\ZZC}^\N$  & $T_{\ZZC}^\B$ &$T_{\QFZ}$ &$T_{\QFZ}^{\max}$ & $T_{\Z}$ & $T_{\MFW}$  & $T_{\MFLH}$  & $T_{\MFP}$ \\
		\midrule
		\multirow{4}[0]{*}{1} & 0.4   & 9.9   & 10.4  & 9.8   & 10.1  & 7.8   & 13.6  & 6.3   & 28.3  & 22.9  & 7.5   & 5.1   & 10.5  & 10.8  & 10.0 \\
		& 0.8   & 38.0  & 38.4  & 36.9  & 45.7  & 31.1  & 44.4  & 26.8  & 63.3  & 57.6  & 33.0  & 21.8  & 38.3  & 39.0  & 36.6 \\
		& 1.2   & 83.8  & 85.2  & 82.2  & 94.4  & 81.2  & 88.7  & 76.6  & 94.4  & 92.5  & 85.4  & 73.2  & 84.4  & 86.3  & 82.7 \\
		& 1.6   & 99.1  & 99.1  & 98.8  & 99.9  & 98.9  & 99.3  & 98.8  & 100.0 & 99.9  & 99.7  & 97.7  & 99.1  & 99.2  & 99.0 \\
		\multirow{4}[0]{*}{2} & 0.4   & 9.3   & 9.3   & 9.2   & 8.9   & 6.4   & 12.7  & 5.3   & 28.3  & 22.5  & 9.5   & 4.0   & 9.5   & 9.1   & 9.2 \\
		& 0.8   & 38.2  & 38.4  & 37.6  & 45.2  & 31.2  & 43.4  & 25.2  & 61.0  & 55.1  & 35.6  & 22.3  & 37.5  & 38.0  & 36.3 \\
		& 1.2   & 82.0  & 83.2  & 80.7  & 94.1  & 78.6  & 86.9  & 74.0  & 94.6  & 93.1  & 86.3  & 69.6  & 81.6  & 82.5  & 79.6 \\
		& 1.6   & 99.1  & 99.2  & 98.8  & 100.0 & 98.9  & 99.6  & 98.2  & 99.9  & 99.9  & 99.6  & 97.6  & 99.3  & 99.3  & 98.8 \\
		\multirow{4}[0]{*}{3} & 0.4   & 9.9   & 9.9   & 9.7   & 8.9   & 7.0   & 13.0  & 4.9   & 30.4  & 23.8  & 9.6   & 4.6   & 9.5   & 9.5   & 9.6 \\
		& 0.8   & 37.9  & 38.5  & 37.2  & 47.0  & 31.5  & 44.8  & 27.7  & 64.8  & 57.8  & 37.6  & 23.5  & 35.8  & 36.3  & 34.9 \\
		& 1.2   & 83.2  & 84.4  & 82.2  & 94.1  & 79.7  & 87.0  & 75.6  & 93.7  & 92.0  & 85.2  & 71.5  & 81.8  & 82.5  & 79.4 \\
		& 1.6   & 98.8  & 99.2  & 98.6  & 100.0 & 98.6  & 99.4  & 97.8  & 99.8  & 99.6  & 99.7  & 96.7  & 98.5  & 99.0  & 97.8 \\
		\bottomrule
	\end{tabular*}
	\label{power.tab1}}
\end{table}%

For the power comparison, it is inappropriate to compare the powers of tests if they are either too conservative or too liberal. Therefore, we focus solely on scenarios with large sample sizes to ensure that the tests under consideration are neither overly conservative nor overly liberal.
The empirical powers (in \%) of all considered tests under homoscedastic case (S1) when  $\bn=\bn_3$ are given in Table~\ref{power.tab1}. Since when $\rho$ is large, the eigenvalues $\lambda_{ir}$ decay more slowly, making the functional samples noisier, we anticipate that the empirical power of a test decreases as $\rho$ increases for a fixed value of $\delta$.  To investigate this, we consider the following values of $\delta$: $\delta=0.1,0.2,0.3,0.4$ when $\rho=0.1$; $\delta=0.2,0.4,0.6,0.8$ when $\rho=0.5$; and $\delta=0.4,0.8,1.2,1.6$ when $\rho=0.9$. As shown in Table~\ref{power.tab1},   the empirical powers of the tests increase as the values of $\delta$ increase, and they also increase as the sample sizes  grow.  Notably, when $\rho=0.1$, $T_{\FLH}^\N,T_{\FLH}^\B,T_{\ZZC}^\N, T_{\ZZC}^\B, T_{\QFZ}$ and $T_{\QFZ}^{\max}$ demonstrate significantly higher powers compared to the other eight tests.  A clear pattern emerges when analyzing the empirical powers of the tests in relation to their empirical sizes, as $T_{\FLH}^\N,T_{\FLH}^\B,T_{\ZZC}^\N, T_{\ZZC}^\B$,  and $T_{\QFZ}$ exhibit higher empirical sizes, as seen in Table~\ref{size.tab1}. As $\rho$ increases, the empirical sizes of $T_{\FLH}^\N$ and $T_{\ZZC}^\N$ decrease, causing their powers to no longer significantly outperform the previous eight tests when $\rho=0.5$ and 0.9. In contrast, $T_{\FLH}^\B,T_{\ZZC}^\B$, and $T_{\QFZ}$ remain liberal, as indicated in Table~\ref{size.tab1}, and thus maintain higher empirical powers than the aforementioned eight tests. Moreover, $T_{\Z}$ consistently demonstrates the least power due to its conservative behavior, as shown in Table~\ref{size.tab1}.  This trend indicates that these competing tests may struggle to maintain effective size control, even with a larger sample size. This highlights the critical importance of size control in evaluating test performance and underscores the need for an optimal method that can reliably manage size to avoid potentially misleading results. Interestingly, $T_{\QFZ}^{\max}$ is the most powerful test when $\rho=0.1$ and 0.5, but becomes comparable to the three new tests when the functional data are less correlated ($\rho=0.9$). The empirical powers of $T_{\MFW},T_{\MFLH}$ and $T_{\MFP}$ are generally similar to those of the tests from  \cite{gorecki2017multivariate} since their empirical sizes in Table~\ref{size.tab1} are generally comparable.  

Next, we evaluate the finite-sample performance of the tests under consideration in the context of the heteroscedastic one-way FMANOVA problem. The empirical sizes of the tests considered under this scenario (S2) are detailed in Table~\ref{size.tab2}. Several important conclusions can be drawn from this table. Firstly, when $\rho=0.1$, $T_{\FLH}^\N,T_{\FLH}^\B,T_{\ZZC}^\N$, $T_{\ZZC}^\B$, and $T_{\QFZ}$ remain relatively liberal, while $T_{\Z}$ continues to exhibit conservative behavior. $T_{\QFZ}^{\max}$ performs well with smaller sample sizes but becomes increasingly conservative as the sample size grows. The tests proposed by \cite{gorecki2017multivariate},  developed under the assumption of homoscedasticity, become conservative and perform less effectively compared to their results in Table~\ref{size.tab1}. In contrast, the three newly proposed tests generally exhibit strong performance.  As the value of $\rho$ increases, the differences between the covariance matrix functions also grow larger. When $\rho=0.9$, the performance of \cite{gorecki2017multivariate}'s tests further deteriorates due to the increasing variance differences between groups, becoming notably more conservative, especially with larger sample sizes. The other six competing tests are either overly conservative or excessively liberal. In contrast, the three modified tests maintain strong performance, demonstrating their robustness under these challenging conditions.
\begin{table}[!h]
	\centering
	\caption{Empirical sizes (in $\%$)  in Simulation 1 under  heteroscedastic case (S2).}
	\vspace{1mm}
	\setlength{\tabcolsep}{1pt} 
	\scalebox{0.9}{
		\begin{tabular*}{\textwidth}{@{\extracolsep{\fill}}cccccccccccccccc@{\extracolsep{\fill}}}
			\toprule
			\multicolumn{3}{l}{$\rho=0.1$} \\
			\midrule
			Model  & $\bn$     & W     & LH    & P     & R     & $T_{\FLH}^\N$  & $T_{\FLH}^\B$  & $T_{\ZZC}^\N$  & $T_{\ZZC}^\B$ &$T_{\QFZ}$ &$T_{\QFZ}^{\max}$ & $T_{\Z}$ & $T_{\MFW}$  & $T_{\MFLH}$  & $T_{\MFP}$ \\
			\midrule
			\multirow{3}[0]{*}{1} & $\bn_1$    & 4.8   & 5.2   & 4.8   & 5.9   & 16.3  & 20.7  & 13.1  & 22.6  & 21.1  & 5.6   & 3.1   & 8.1   & 8.6   & 6.3 \\
			& $\bn_2$    & 4.3   & 4.6   & 3.9   & 4.5   & 11.6  & 13.7  & 10.4  & 16.8  & 14.7  & 5.3   & 4.3   & 6.9   & 7.1   & 5.5 \\
			& $\bn_3$    & 3.9   & 4.0   & 3.5   & 3.9   & 7.0   & 8.2   & 8.2   & 11.6  & 8.0   & 3.0   & 5.5   & 6.5   & 6.6   & 5.7 \\
			\multirow{3}[0]{*}{2} & $\bn_1$    & 4.2   & 4.7   & 4.4   & 5.1   & 16.5  & 22.0  & 12.3  & 24.7  & 21.7  & 3.8   & 1.8   & 6.4   & 6.7   & 4.6 \\
			& $\bn_2$    & 4.1   & 4.2   & 3.9   & 4.5   & 12.9  & 15.7  & 10.6  & 17.6  & 15.4  & 5.0   & 3.6   & 5.7   & 5.7   & 5.0 \\
			&$\bn_3$    & 2.1   & 2.2   & 2.2   & 3.1   & 6.2   & 7.4   & 6.9   & 10.0  & 7.1   & 3.6   & 3.7   & 4.9   & 4.7   & 4.4 \\
			\multirow{3}[0]{*}{3} & $\bn_1$    & 4.2   & 4.5   & 4.3   & 5.2   & 16.9  & 21.1  & 12.4  & 24.7  & 22.5  & 4.8   & 2.7   & 5.2   & 5.3   & 4.4 \\
			& $\bn_2$    & 3.5   & 3.7   & 3.7   & 4.0   & 10.2  & 13.3  & 8.3   & 16.0  & 13.1  & 4.2   & 2.3   & 4.5   & 4.5   & 3.2 \\
			& $\bn_3$    & 2.9   & 2.6   & 2.8   & 4.2   & 6.0   & 6.9   & 6.2   & 10.3  & 6.7   & 3.6   & 3.5   & 4.0   & 4.3   & 3.7 \\
			\midrule
			\multicolumn{2}{c}{ARE} & 24.44 & 21.56 & 25.56 & 15.56 & 130.22 & 186.67 & 96.44 & 242.89 & 189.56 & 17.56 & 34.44 & 23.11 & 25.56 & 16.00 \\
			\midrule
			\multicolumn{3}{l}{$\rho=0.5$} \\
			\midrule
			Model  & $\bn$     & W     & LH    & P     & R     & $T_{\FLH}^\N$  & $T_{\FLH}^\B$  & $T_{\ZZC}^\N$  & $T_{\ZZC}^\B$ &$T_{\QFZ}$ &$T_{\QFZ}^{\max}$ & $T_{\Z}$ & $T_{\MFW}$  & $T_{\MFLH}$  & $T_{\MFP}$ \\
			\midrule
			\multirow{3}[0]{*}{1} & $\bn_1$    & 4.3   & 4.4   & 4.0   & 5.6   & 7.2   & 13.6  & 11.8  & 36.6  & 29.5  & 5.9   & 2.3   & 7.2   & 7.6   & 5.7 \\
			& $\bn_2$    & 3.7   & 3.5   & 3.4   & 3.6   & 5.6   & 8.4   & 9.4   & 23.2  & 16.9  & 4.3   & 3.7   & 6.6   & 6.9   & 6.0 \\
			& $\bn_3$    & 1.8   & 1.9   & 1.8   & 3.6   & 2.9   & 4.7   & 6.8   & 15.3  & 7.4   & 2.6   & 4.1   & 6.2   & 5.9   & 5.6 \\
			\multirow{3}[0]{*}{2} & $\bn_1$    & 4.6   & 4.4   & 4.3   & 5.3   & 6.6   & 13.4  & 10.6  & 36.5  & 27.8  & 5.1   & 1.1   & 6.3   & 6.5   & 5.4 \\
			& $\bn_2$    & 3.3   & 3.2   & 3.1   & 5.1   & 5.4   & 9.1   & 9.5   & 22.8  & 17.1  & 4.6   & 2.7   & 6.1   & 6.6   & 5.5 \\
			& $\bn_3$    & 2.3   & 2.1   & 2.0   & 3.1   & 3.0   & 4.3   & 5.3   & 12.4  & 6.9   & 2.8   & 3.8   & 5.1   & 5.1   & 4.6 \\
			\multirow{3}[0]{*}{3} & $\bn_1$    & 3.0   & 2.9   & 3.0   & 4.1   & 5.5   & 12.6  & 9.8   & 35.3  & 26.8  & 4.9   & 1.6   & 4.3   & 4.9   & 3.5 \\
			& $\bn_2$    & 3.5   & 3.6   & 3.4   & 3.7   & 4.3   & 7.1   & 6.2   & 22.3  & 14.4  & 4.1   & 2.1   & 4.5   & 4.9   & 3.9 \\
			& $\bn_3$    & 2.2   & 2.2   & 2.2   & 3.6   & 2.8   & 4.6   & 5.4   & 14.0  & 6.9   & 3.9   & 4.0   & 4.5   & 5.0   & 4.3 \\
			\midrule
			\multicolumn{2}{c}{ARE} & 36.22 & 37.33 & 39.56 & 20.67 & 27.33 & 79.11 & 66.22 & 385.33 & 241.56 & 19.56 & 43.56 & 20.44 & 19.56 & 15.33 \\
			\midrule
			\multicolumn{3}{l}{$\rho=0.9$} \\
			\midrule
			Model  & $\bn$     & W     & LH    & P     & R    & $T_{\FLH}^\N$  & $T_{\FLH}^\B$  & $T_{\ZZC}^\N$  & $T_{\ZZC}^\B$ &$T_{\QFZ}$ &$T_{\QFZ}^{\max}$ & $T_{\Z}$ & $T_{\MFW}$  & $T_{\MFLH}$  & $T_{\MFP}$ \\
			\midrule
			\multirow{3}[0]{*}{1} & $\bn_1$    & 3.3   & 3.5   & 3.3   & 5.4   & 1.4   & 7.4   & 7.4   & 53.2  & 34.3  & 6.2   & 0.5   & 6.6   & 6.9   & 6.4 \\
			& $\bn_2$    & 2.2   & 2.2   & 2.2   & 3.3   & 1.2   & 4.2   & 4.6   & 35.6  & 17.7  & 5.1   & 1.6   & 5.1   & 5.1   & 4.7 \\
			& $\bn_3$    & 1.1   & 1.0   & 1.0   & 2.3   & 0.6   & 1.7   & 3.3   & 17.1  & 5.5   & 3.6   & 2.3   & 5.3   & 5.5   & 5.3 \\
			\multirow{3}[0]{*}{2} & $\bn_1$    & 3.6   & 3.7   & 3.5   & 4.1   & 1.6   & 7.5   & 6.3   & 52.6  & 32.8  & 5.2   & 0.4   & 5.6   & 5.8   & 5.3 \\
			& $\bn_2$    & 2.6   & 2.8   & 2.5   & 4.0   & 1.0   & 4.2   & 4.3   & 32.2  & 17.8  & 6.1   & 1.5   & 4.8   & 4.8   & 4.7 \\
			& $\bn_3$    & 1.4   & 1.3   & 1.3   & 3.1   & 1.1   & 2.1   & 3.6   & 17.4  & 5.6   & 3.6   & 2.8   & 5.0   & 5.4   & 4.9 \\
			\multirow{3}[0]{*}{3} & $\bn_1$    & 2.1   & 2.1   & 2.2   & 3.5   & 0.8   & 6.0   & 4.5   & 49.5  & 29.7  & 5.1   & 0.1   & 3.6   & 3.6   & 3.2 \\
			& $\bn_2$    & 2.8   & 2.9   & 2.7   & 3.3   & 1.5   & 5.5   & 4.7   & 31.4  & 15.9  & 4.5   & 1.3   & 5.3   & 5.2   & 5.3 \\
			& $\bn_3$    & 1.1   & 1.1   & 1.1   & 2.5   & 0.3   & 1.9   & 2.9   & 17.3  & 5.9   & 4.4   & 1.6   & 4.9   & 5.1   & 4.9 \\
			\midrule
			\multicolumn{2}{c}{ARE} & 55.11 & 54.22 & 56.00 & 31.78 & 78.89 & 38.44 & 24.00 & 580.67 & 267.11 & 14.67 & 73.11 & 10.22 & 12.44 & 10.89 \\
			\bottomrule
		\end{tabular*}
		\label{size.tab2}}
\end{table}%

\begin{table}[!h]
	\centering
	\caption{Empirical powers (in $\%$)  in Simulation 1 under heteroscedastic  case (S2)  when $\bn=\bn_3$.}
	\vspace{1mm}
	\setlength{\tabcolsep}{1pt} 
	\scalebox{0.9}{
	\begin{tabular*}{\textwidth}{@{\extracolsep{\fill}}cccccccccccccccc@{\extracolsep{\fill}}}
		\toprule
		\multicolumn{3}{l}{$\rho=0.1$}\\
		\midrule
	Model	& $\delta$     & W     & LH    & P     & R     & $T_{\FLH}^\N$  & $T_{\FLH}^\B$  & $T_{\ZZC}^\N$  & $T_{\ZZC}^\B$ &$T_{\QFZ}$ &$T_{\QFZ}^{\max}$ & $T_{\Z}$ & $T_{\MFW}$  & $T_{\MFLH}$  & $T_{\MFP}$ \\
		\midrule
		\multirow{4}[0]{*}{1} & 0.1   & 5.4   & 5.7   & 5.6   & 5.9   & 10.4  & 12.8  & 12.9  & 17.4  & 12.6  & 12.1  & 7.7   & 9.4   & 9.3   & 8.6 \\
		& 0.2   & 18.5  & 18.0  & 18.2  & 17.1  & 28.1  & 30.2  & 30.7  & 39.0  & 30.6  & 75.5  & 22.5  & 25.2  & 26.1  & 24.6 \\
		& 0.3   & 49.6  & 50.3  & 48.4  & 51.8  & 63.5  & 67.3  & 65.9  & 74.2  & 66.0  & 100.0 & 54.0  & 58.7  & 59.2  & 55.9 \\
		& 0.4   & 85.9  & 86.6  & 84.0  & 89.2  & 92.8  & 94.3  & 93.8  & 95.9  & 93.9  & 100.0 & 90.2  & 90.7  & 91.1  & 89.1 \\
		\multirow{4}[0]{*}{2} & 0.1   & 5.5   & 5.0   & 5.6   & 4.5   & 9.9   & 10.9  & 10.5  & 15.0  & 10.6  & 10.8  & 6.4   & 7.8   & 7.6   & 7.6 \\
		& 0.2   & 16.1  & 16.1  & 16.2  & 14.1  & 25.9  & 28.6  & 27.2  & 35.2  & 28    & 74.6  & 19.0  & 22.1  & 22.6  & 20.6 \\
		& 0.3   & 48.6  & 49.6  & 47.4  & 50.3  & 61.8  & 66.1  & 64.4  & 73.4  & 65.9  & 99.3  & 53.2  & 57.1  & 57.5  & 54.2 \\
		& 0.4   & 83.5  & 84.5  & 82.0  & 87.3  & 91.2  & 92.9  & 92.2  & 95.5  & 92.3  & 100   & 87.4  & 88.2  & 89.5  & 86.3 \\
		\multirow{4}[0]{*}{3} & 0.1   & 4.8   & 4.8   & 5.0   & 5.2   & 8.6   & 10.3  & 9.3   & 14.8  & 10.0  & 10.8  & 5.3   & 6.5   & 7.0   & 6.1 \\
		& 0.2   & 15.6  & 15.7  & 15.7  & 16.1  & 27.4  & 31.9  & 29.5  & 38.9  & 30.6  & 76.4  & 19.9  & 22.4  & 22.5  & 21.3 \\
		& 0.3   & 49.7  & 50.6  & 48.4  & 52.8  & 64.6  & 68.4  & 66.2  & 76.0  & 67.5  & 99.6  & 54.6  & 56.1  & 57.4  & 53.4 \\
		& 0.4   & 85.9  & 87.6  & 84.8  & 89.1  & 92.2  & 93.5  & 93.1  & 96.1  & 93.4  & 100   & 88.0  & 88.0  & 89.1  & 86.4 \\
		\midrule
		\multicolumn{3}{l}{$\rho=0.5$} \\
		\midrule
	Model	& $\delta$     & W     & LH    & P     & R     & $T_{\FLH}^\N$  & $T_{\FLH}^\B$  & $T_{\ZZC}^\N$  & $T_{\ZZC}^\B$ &$T_{\QFZ}$ &$T_{\QFZ}^{\max}$ & $T_{\Z}$ & $T_{\MFW}$  & $T_{\MFLH}$  & $T_{\MFP}$ \\
		\midrule
		\multirow{4}[0]{*}{1} & 0.2   & 3.9   & 3.7   & 3.9   & 5.2   & 5.0   & 7.5   & 9.5   & 18.6  & 10.7  & 5.1   & 7.1   & 8.9   & 9.1   & 8.4 \\
		& 0.4   & 9.8   & 10.3  & 9.6   & 12.2  & 12.6  & 16.5  & 20.0  & 33.7  & 22.5  & 21.1  & 15.8  & 19.6  & 19.8  & 18.2 \\
		& 0.6   & 28.6  & 30.0  & 27.9  & 38.1  & 32.6  & 39.7  & 46.2  & 63.7  & 49.9  & 70.7  & 39.7  & 43.8  & 45.6  & 42.2 \\
		& 0.8   & 64.1  & 66.4  & 61.4  & 75.9  & 69.5  & 75.1  & 79.4  & 88.1  & 81.1  & 97.0  & 74.2  & 78.0  & 79.4  & 74.7 \\
		\multirow{4}[0]{*}{2} & 0.2   & 3.6   & 3.7   & 3.7   & 4.9   & 4.1   & 6.0   & 7.6   & 16.7  & 9.2   & 5.7   & 5.5   & 7.5   & 7.6   & 6.6 \\
		& 0.4   & 9.1   & 9.6   & 8.8   & 10.4  & 11.1  & 14.2  & 18.0  & 33.0  & 20.6  & 20.2  & 13.4  & 16.1  & 16.6  & 15.0 \\
		& 0.6   & 29.6  & 30.9  & 28.0  & 35.6  & 33.3  & 39.7  & 41.9  & 61.0  & 46.8  & 69.1  & 35.7  & 40.7  & 42.0  & 39.2 \\
		& 0.8   & 60.1  & 62.7  & 58.9  & 72.8  & 66.7  & 71.9  & 77.0  & 87.8  & 80.2  & 97.3  & 71.9  & 74.0  & 75.0  & 71.0 \\
		\multirow{4}[0]{*}{3} & 0.2   & 3.3   & 3.4   & 3.1   & 4.2   & 4.1   & 6.2   & 7.8   & 16.9  & 10.0  & 5.9   & 5.4   & 7.2   & 7.6   & 6.6 \\
		& 0.4   & 9.8   & 9.8   & 9.9   & 10.3  & 11.5  & 16.5  & 17.0  & 34.9  & 21.5  & 23.4  & 13.4  & 17.2  & 17.9  & 16.2 \\
		& 0.6   & 29.1  & 30.2  & 28.3  & 37.2  & 34.2  & 40.8  & 45.6  & 65.0  & 49.9  & 70.7  & 38.3  & 41.5  & 42.7  & 39.0 \\
		& 0.8   & 62.8  & 65.2  & 60.1  & 76.3  & 68.5  & 74.4  & 78.3  & 89.4  & 81.2  & 97.8  & 72.0  & 74.1  & 75.7  & 70.6 \\
		\midrule
		\multicolumn{3}{l}{$\rho=0.9$}\\
		\midrule
	Model & $\delta$      & W     & LH    & P     & R     & $T_{\FLH}^\N$  & $T_{\FLH}^\B$  & $T_{\ZZC}^\N$  & $T_{\ZZC}^\B$ &$T_{\QFZ}$ &$T_{\QFZ}^{\max}$ & $T_{\Z}$ & $T_{\MFW}$  & $T_{\MFLH}$  & $T_{\MFP}$ \\
		\midrule
		\multirow{4}[0]{*}{1} & 0.4   & 2.8   & 2.6   & 2.7   & 4.9   & 1.2   & 3.7   & 5.6   & 23.9  & 8.5   & 4.7   & 4.3   & 8.9   & 8.8   & 8.4 \\
		& 0.8   & 9.6   & 9.9   & 9.5   & 17.3  & 7.5   & 13.1  & 16.4  & 48.1  & 23.3  & 12.7  & 13.0  & 23.2  & 23.7  & 22.9 \\
		& 1.2   & 36.9  & 39.0  & 34.9  & 64.7  & 31.3  & 45.7  & 50.8  & 81.5  & 59.6  & 48.7  & 45.6  & 58.4  & 59.8  & 57.0 \\
		& 1.6   & 81.1  & 82.4  & 78.2  & 96.6  & 77.8  & 86.2  & 87.7  & 97.7  & 91.2  & 90.8  & 85.0  & 91.8  & 92.4  & 90.6 \\
		\multirow{4}[0]{*}{2} & 0.4   & 3.3   & 3.2   & 3.2   & 4.1   & 1.7   & 3.9   & 5.8   & 23.1  & 8.8   & 5.6   & 4.4   & 7.5   & 7.6   & 7.3 \\
		& 0.8   & 10.3  & 10.7  & 9.9   & 17.3  & 7.2   & 13.0  & 16.8  & 45.1  & 22.3  & 13.7  & 14.6  & 21.9  & 22.3  & 21.3 \\
		& 1.2   & 37.1  & 38.3  & 36.1  & 62.8  & 30.8  & 44.9  & 48.1  & 78.8  & 56.3  & 50.9  & 43.7  & 57.7  & 58.7  & 55.6 \\
		& 1.6   & 78.3  & 81.3  & 76.4  & 96.5  & 74.4  & 84.7  & 86.6  & 97.4  & 91.1  & 89.9  & 82.7  & 89.4  & 90.4  & 88.5 \\
		\multirow{4}[0]{*}{3} & 0.4   & 2.3   & 2.5   & 2.3   & 3.7   & 1.0   & 3.9   & 5.2   & 25.0  & 8.6   & 5.9   & 3.5   & 8.4   & 8.5   & 8.3 \\
		& 0.8   & 9.8   & 10.2  & 9.6   & 18.7  & 6.2   & 14.1  & 17.3  & 49.0  & 25.3  & 15.5  & 13.5  & 23.3  & 23.2  & 22.4 \\
		& 1.2   & 38.1  & 39.4  & 36.8  & 65.1  & 32.2  & 47.0  & 50.1  & 82.7  & 60.6  & 52.5  & 45.0  & 57.9  & 59.3  & 55.8 \\
		& 1.6   & 80.4  & 81.7  & 77.9  & 97.0  & 76.3  & 85.3  & 87.0  & 96.9  & 89.6  & 89.6  & 83.3  & 89.3  & 91.1  & 88.5 \\
		\bottomrule
	\end{tabular*}
	\label{power.tab2}}
\end{table}%

Table~\ref{power.tab2} presents the empirical powers (in \%) of all considered tests under the heteroscedastic case (S2) when  $\bn=\bn_3$. The findings are consistent with those observed in Table~\ref{power.tab1}. Notably, $T_{\QFZ}^{\max}$ remains the most powerful test when $\rho=0.1$, but its empirical powers declines as $\rho$ increases, consistent with its empirical sizes shown in Table~\ref{size.tab2}.  $T_{\FLH}^\N,T_{\FLH}^\B,T_{\ZZC}^\N$, $T_{\ZZC}^\B$, and $T_{\QFZ}$  exhibit higher powers in the second batch due to their empirical sizes from Table~\ref{size.tab2} being significantly greater than  5\% when $\rho=0.1$. Additionally, the empirical powers of the three newly proposed tests, i.e.,  $T_{\MFW}$, $T_{\MFLH}$ and $T_{\MFP}$, are either larger than or comparable to those of the tests from \cite{gorecki2017multivariate} and \cite{zhu2023general}, which are notably conservative in Table~\ref{size.tab2}. Overall, these results highlight the efficacy of the proposed tests, especially in maintaining robust power levels in the face of heteroscedasticity.

\subsection{Simulation 2: Two-sample problem}
When the one-way FMANOVA (\ref{BFH0.sec2}) is rejected, it is reasonable to proceed with pairwise comparisons.  For example, we can compare $\bfeta_1(t)$ and $\bfeta_4(t)$, specifically testing $\H_0:\bfeta_1(t)=\bfeta_4(t)$ by setting $\bC=(1,0,0,-1)$ and  $\bC_0(t)=\bm{0}$  for all $t\in\calT$.  This results in a two-sample problem previously explored  by \cite{Qiu2021}. Consequently, we can assess the finite-sample performance of $T_{\MFW}$, $T_{\MFLH}$ and $T_{\MFP}$ in comparison to \cite{Qiu2021}'s two-sample tests based on the integration and supremum of the pointwise Hotelling's $T^2$-test statistics, denoted as $T_{\QCZ}$  and $T_{\QCZ}^{\max}$, respectively. Given that $T_{\FLH}^\N,T_{\FLH}^\B,T_{\ZZC}^\N,T_{\ZZC}^\B$, and $T_{\Z}$  do not perform well in the aforementioned one-way FMANOVA context for small sample sizes, and $T_{\QFZ}$ and $T_{\QFZ}^{\max}$ behave similarly to $T_{\QCZ}$  and $T_{\QCZ}^{\max}$ for the two-sample problem,  we will focus on comparing $T_{\MFW}$, $T_{\MFLH}$ and $T_{\MFP}$ with W, LH, P, R, $T_{\QCZ}$  and $T_{\QCZ}^{\max}$ for this two-sample scenario. 
\begin{table}[!h]
	\centering
	\caption{Empirical sizes (in $\%$)  in Simulation 2 under heteroscedastic  case (S2) when $\bC=(1,0,0,-1)$.}
	\vspace{1mm}
	\setlength{\tabcolsep}{1pt} 
	\scalebox{0.9}{
	\begin{tabular*}{\textwidth}{@{\extracolsep{\fill}}cccccccccccccc@{\extracolsep{\fill}}}
		\toprule
		$\rho=0.1$ \\
		\midrule
		Model  & $\bn$     & W     & LH    & P     & R     & $T_{\QCZ}$  & $T_{\QCZ}^{\max}$    & $T_{\MFW}$  & $T_{\MFLH}$  & $T_{\MFP}$ \\
		\midrule
		\multirow{3}[0]{*}{1} & $\bn_1$    & 3.3   & 3.3   & 3.3   & 3.7   & 29.2  & 4.4   & 6.3   & 7.5   & 4.4 \\
		& $\bn_2$    & 3.7   & 3.9   & 3.8   & 4.2   & 14.6  & 4.8   & 7.4   & 7.9   & 6.4 \\
		& $\bn_3$    & 1.9   & 1.9   & 1.8   & 1.8   & 7.3   & 2.7   & 4.9   & 5.0   & 4.1 \\
		\multirow{3}[0]{*}{2} & $\bn_1$    & 4.5   & 4.5   & 4.5   & 4.5   & 29.2  & 4.8   & 6.6   & 7.0   & 5.2 \\
		& $\bn_2$    & 3.4   & 3.6   & 3.4   & 3.6   & 15.7  & 4.2   & 6.1   & 6.9   & 4.9 \\
		& $\bn_3$    & 2.3   & 2.6   & 2.1   & 3.0   & 8.2   & 3.3   & 5.5   & 5.8   & 4.9 \\
		\multirow{3}[0]{*}{3} & $\bn_1$    & 5.9   & 5.4   & 5.6   & 5.4   & 29.3  & 5.8   & 7.5   & 8.5   & 5.5 \\
		& $\bn_2$    & 3.6   & 3.8   & 3.4   & 3.9   & 15.3  & 3.8   & 4.8   & 5.2   & 3.9 \\
		& $\bn_3$    & 2.2   & 2.4   & 2.1   & 2.6   & 6.8   & 2.4   & 4.0   & 4.4   & 3.6 \\
		\midrule
		\multicolumn{2}{c}{ARE} & 35.56 & 32.00 & 36.00 & 29.11 & 245.78 & 23.11 & 23.78 & 32.00 & 14.00 \\
		\midrule
		$\rho=0.5$ \\
		\midrule
		Model & $\bn$     & W     & LH    & P     & R     & $T_{\QCZ}$  & $T_{\QCZ}^{\max}$     & $T_{\MFW}$  & $T_{\MFLH}$  & $T_{\MFP}$ \\
		\midrule
		\multirow{3}[0]{*}{1} & $\bn_1$    & 3.2   & 3.1   & 3.2   & 2.8   & 39.2  & 4.7   & 6.6   & 7.6   & 5.0 \\
		& $\bn_2$    & 2.3   & 2.4   & 2.2   & 2.8   & 15.3  & 5.2   & 6.2   & 6.7   & 5.4 \\
		& $\bn_3$    & 1.0   & 1.1   & 0.9   & 1.5   & 5.1   & 2.8   & 3.7   & 4.2   & 3.3 \\
		\multirow{3}[0]{*}{2} & $\bn_1$    & 3.9   & 3.9   & 4.7   & 3.2   & 37.1  & 6.1   & 7.1   & 7.1   & 5.7 \\
		& $\bn_2$    & 2.1   & 2.3   & 2.1   & 2.9   & 16.8  & 3.8   & 5.3   & 6.0   & 5.2 \\
		& $\bn_3$    & 1.6   & 1.4   & 1.5   & 2.3   & 7.0   & 3.8   & 6.2   & 6.8   & 5.4 \\
		\multirow{3}[0]{*}{3} & $\bn_1$    & 3.3   & 3.4   & 3.1   & 3.8   & 36.2  & 6.3   & 5.2   & 6.1   & 4.3 \\
		& $\bn_2$    & 2.0   & 2.0   & 2.2   & 3.0   & 18.4  & 4.0   & 4.8   & 5.2   & 4.7 \\
		& $\bn_3$    & 1.8   & 1.8   & 1.7   & 2.3   & 5.9   & 2.8   & 5.3   & 5.5   & 4.8 \\
		\midrule
		\multicolumn{2}{c}{ARE} & 52.89 & 52.44 & 52.00 & 45.33 & 302.22 & 23.78 & 18.67 & 26.22 & 10.22 \\
		\midrule
		$\rho=0.9$ \\
		\midrule
		Model	 & $\bn$     & W     & LH    & P     & R     & $T_{\QCZ}$  & $T_{\QCZ}^{\max}$     & $T_{\MFW}$  & $T_{\MFLH}$  & $T_{\MFP}$ \\
		\midrule
		\multirow{3}[0]{*}{1} & $\bn_1$    & 2.4   & 2.8   & 2.4   & 3.6   & 43.5  & 4.4   & 6.8   & 6.9   & 6.1 \\
		& $\bn_2$    & 1.5   & 1.5   & 1.5   & 1.9   & 15.4  & 4.6   & 6.2   & 6.5   & 5.6 \\
		& $\bn_3$    & 0.3   & 0.2   & 0.4   & 1.1   & 2.5   & 2.8   & 3.7   & 3.7   & 3.6 \\
		\multirow{3}[0]{*}{2} & $\bn_1$    & 2.2   & 2.3   & 2.2   & 2.6   & 46.4  & 6.3   & 6.3   & 6.2   & 6.0 \\
		& $\bn_2$    & 0.8   & 0.8   & 0.8   & 2.1   & 16.3  & 3.7   & 5.2   & 5.4   & 4.9 \\
		& $\bn_3$    & 0.5   & 0.6   & 0.5   & 1.5   & 4.0   & 2.5   & 4.9   & 5.2   & 4.9 \\
		\multirow{3}[0]{*}{3} & $\bn_1$    & 2.8   & 2.7   & 2.8   & 3.7   & 41.2  & 5.7   & 5.1   & 5.1   & 4.5 \\
		& $\bn_2$    & 1.0   & 1.1   & 0.9   & 2.0   & 16.5  & 4.7   & 6.1   & 6.1   & 5.5 \\
		& $\bn_3$    & 0.9   & 0.9   & 1.0   & 1.8   & 4.8   & 3.0   & 6.0   & 6.0   & 5.8 \\
		\midrule
		\multicolumn{2}{c}{ARE} & 72.44 & 71.33 & 72.22 & 54.89 & 340.00 & 25.11 & 18.00 & 19.33 & 13.56 \\
		\bottomrule
	\end{tabular*}}
	\label{size.tab3}%
\end{table}%

Table~\ref{size.tab3} summarizes the empirical sizes of all considered tests in Simulation 2 under the heteroscedastic case (S2) with $\bC=(1,0,0,-1)$. We can conclude that, among the three proposed tests,  $T_{\MFP}$ demonstrates the best performance regarding  ARE values, achieving the smallest ARE values of 14.00, 10.22, and 13.56 for $\rho=0.1, 0.5$, and $0.9$, respectively. $T_{\MFW}$ and  $T_{\MFLH}$ also perform well, with empirical sizes of $T_{\MFW}$  ranging from 4.0\% to 7.5\% for $\rho=0.1$, 3.7\% to 7.1\% for $\rho=0.5$, and 3.7\% to 6.8\% for $\rho=0.9$. In contrast, the empirical sizes of $T_{\MFLH}$ range from 4.4\% to 8.5\% for $\rho=0.1$, 4.2\% to 7.6\% for $\rho=0.5$, and 3.7\% to 6.9\% for $\rho=0.9$, respectively. While $T_{\QCZ}$ exhibits a liberal behavior and only achieves acceptable size control when $\bn=\bn_3$. $T_{\QCZ}^{\max}$ becomes conservative with larger sample sizes. Additionally, the four permutation-based tests proposed by \cite{gorecki2017multivariate} show conservative performance consistent with the results in Table~\ref{size.tab2}. To save space, we do not present the empirical powers of these tests in two-sample problem since the conclusions drawn from them are similar to those drawn from Table~\ref{power.tab1} and Table~\ref{power.tab2}.

\subsection{Simulation 3: Testing linear combinations of mean functions}
It is important to note that, under this simulation setting, the null hypothesis $\H_0:\bfeta_1(t)=\bfeta_4(t)$ is equivalent to the  specific linear hypothesis $\H_0:\bfeta_1(t)-3\bfeta_2(t)+2\bfeta_4(t)=\bm{0}$. Consequently, it is reasonable to further assess the finite-sample performance of  $T_{\MFW}$, $T_{\MFLH}$ and $T_{\MFP}$  in comparison to the test proposed by \cite{zhu2023general} by setting  $\bC=(1,-3,0,2)$ and  $\bC_0(t)=\bm{0}$ for all $t\in\calT$. The empirical sizes (in \%) of $T_{\Z}$, $T_{\MFW}$, $T_{\MFLH}$ and $T_{\MFP}$  under heteroscedastic case (S2) when $\bC=(1,-3,0,2)$ are displayed in Table~\ref{size.tab4}. It is observed that the empirical sizes of $T_{\Z}$ are not as small as those in Table~\ref{size.tab1}, indicating the test to become somewhat liberal.  This raises questions about the effectiveness of the adjustment coefficient proposed in \cite{zhu2023general}.  $T_{\MFP}$ demonstrates superior performance over $T_{\MFW}$ and $T_{\MFLH}$ when $\rho=0.1$ and 0.5, as  $T_{\MFW}$ and $T_{\MFLH}$  exhibit slight liberal behavior with small sample sizes. All three tests perform well when the functional data are nearly uncorrelated.

\begin{table}[!h]
	\centering
	\caption{Empirical sizes (in $\%$)  under heteroscedastic  case (S2) when $\bC=(1,-3,0,2)$.}
	\vspace{1mm}
	\setlength{\tabcolsep}{1pt} 
	\scalebox{0.9}{
	\begin{tabular*}{\textwidth}{@{\extracolsep{\fill}}cccccccccccccc@{\extracolsep{\fill}}}
		\toprule
		&       & \multicolumn{4}{c}{$\rho=0.1$}   & \multicolumn{4}{c}{$\rho=0.5$}   & \multicolumn{4}{c}{$\rho=0.9$} \\
		\midrule
		Model  & $\bn$        & $T_{\Z}$ & $T_{\MFW}$  & $T_{\MFLH}$  & $T_{\MFP}$     & $T_{\Z}$ & $T_{\MFW}$  & $T_{\MFLH}$  & $T_{\MFP}$    & $T_{\Z}$ & $T_{\MFW}$  & $T_{\MFLH}$  & $T_{\MFP}$ \\
		\midrule
		\multirow{3}[0]{*}{1} & $\bn_1$    & 9.9  & 9.3   & 10.3  & 7.4   & 10.1  & 7.5   & 8.6   & 5.8   & 7.3  & 6.0   & 6.1   & 5.3 \\
		& $\bn_2$    & 8.0  & 7.4   & 7.9   & 6.4   & 8.9  & 6.3   & 6.5   & 5.6   & 7.6  & 6.4   & 6.5   & 6.0 \\
		& $\bn_3$    & 9.4  & 6.7   & 7.2   & 6.6   & 9.8  & 6.5   & 6.6   & 5.5   & 7.7  & 5.3   & 5.5   & 5.0 \\
		\multirow{3}[0]{*}{2} & $\bn_1$    & 7.2  & 7.1   & 7.7   & 6.0   & 8.7  & 6.7   & 7.1   & 5.8   & 7.0  & 5.4   & 5.6   & 4.7 \\
		& $\bn_2$    & 7.8  & 6.3   & 6.8   & 5.2   & 9.1  & 6.1   & 6.9   & 4.6   & 7.3  & 4.1   & 4.2   & 3.8 \\
		& $\bn_3$    & 8.2  & 6.2   & 6.8   & 5.9   & 9.5  & 5.5   & 5.7   & 5.0   & 7.5  & 4.6   & 4.6   & 4.2 \\
		\multirow{3}[0]{*}{3} & $\bn_1$    & 8.4  & 7.7   & 8.2   & 6.3   & 9.5  & 7.1   & 7.7   & 6.0   & 5.1  & 5.2   & 4.9   & 4.7 \\
		& $\bn_2$    & 6.2  & 4.4   & 5.1   & 3.9   & 8.2  & 4.9   & 5.2   & 4.4   & 8.2  & 4.9   & 4.7   & 4.3 \\
		& $\bn_3$    & 7.3  & 5.3   & 5.6   & 5.1   & 7.8  & 4.7   & 5.0   & 4.7   & 6.3  & 4.6   & 4.7   & 4.2 \\
		\midrule
		\multicolumn{2}{c}{ARE} & 60.89 & 36.89 & 45.78 & 22.22 & 81.33 & 24.67 & 31.78 & 11.11 & 42.22 & 11.33 & 12.44 & 12.00 \\
		\bottomrule
	\end{tabular*}}
	\label{size.tab4}%
\end{table}%

In conclusion, we conducted a series of hypothesis tests in this section by varying the coefficient matrix $\bC$. The results demonstrate that the three proposed tests, $T_{\MFW}$, $T_{\MFLH}$ and $T_{\MFP}$, generally perform reasonably well  in both  homoscedastic and heteroscedastic  cases in terms of controlling the Type I error rate,  with clear advantages over existing competitors in the heteroscedastic case.  Among these three tests, $T_{\MFP}$ performs the best overall, though it tends to be slightly conservative for heavy-tailed distributions (Model 3). Therefore, we recommend using $T_{\MFP}$  for light-tailed distributions and $T_{\MFW}$ or $T_{\MFLH}$  for heavy-tailed distributions.

\section{Real Data Applications}\label{real.sec}
As outlined in Section~\ref{intro.sec}, it is appropriate to apply the concept of MFD to the dataset from \cite{soh2023regularised}'s study. This dataset comprises 344 spectra from 43 bottles across seven brands, labeled Brand I through Brand VII. In this section, we assume that the spectra obtained from drops of the same brand are i.i.d. Accordingly, we consider the spectra from 5 bottles of Brand I to represent spectral observations from Italy, those from 8 bottles of Brand II to represent observations from Spain, and those from 8 bottles of Brand III to represent observations from Greece. For each bottle, four drops were sampled using a dropper, resulting in $n_1 = 20$ observations for Italy, $n_2 = 32$ for Spain, and $n_3 = 32$ for Greece, with each observation consisting of $p = 2$ curves corresponding to two scans of each drop. The primary objective is to assess whether the mean spectral observations differ among these three regions, which constitutes a one-way FMANOVA problem. In this section, we apply $T_{\MFW}$, $T_{\MFLH}$, and $T_{\MFP}$, along with their competitors considered in Simulation 1, to the aforementioned dataset. The $p$-values obtained from these tests are exceptionally low, essentially 0, and much smaller than the selected significance level of $\alpha = 5\%$. As a result, we firmly reject the null hypothesis.

Besides the testing results,  we also compare the computational costs for this one-way FMANOVA problem, which was executed on a 16-inch MacBook Pro with 12 cores, 32GB of RAM, Apple M2 Max (version 4.3.1). Note that the implementations of the newly introduced tests are based on the C++ implementations of the most computationally intensive parts, and the test results are obtained simultaneously. Therefore, we report the total execution time for $T_{\MFW}$, $T_{\MFLH}$, and $T_{\MFP}$ together in Table~\ref{real.tab1}. The total execution time (in seconds) for the remaining tests considered is also shown in Table~\ref{real.tab1}. Specifically, we use the function \texttt{fmanova.ptbfr} from the \textsf{R} package \texttt{fdANOVA} to compute the results for W, LH, P, and R, which are returned simultaneously. As a result, we report the computational time for all four tests together. Similarly, since $T_{\FLH}^\N$ and $T_{\FLH}^\B$ share the same test statistic but use different approximation methods, their results are also obtained simultaneously. Consequently, we report the computational time for both $T_{\FLH}^\N$ and $T_{\FLH}^\B$ together, as shown in Table~\ref{real.tab1}. The same approach is applied to $T_{\ZZC}^\N$ and $T_{\ZZC}^\B$. It is seen from Table~\ref{real.tab1}, the permutation-based tests, namely, W, LH, P, and R are indeed time-consuming. In contrast, our proposed tests are significantly faster than both the permutation-based tests (W, LH, P, R) and the bootstrap-based test ($T_{\QFZ}^{\max}$).
\begin{table}[!h]
		\centering
	\setlength{\tabcolsep}{1pt} 
	\caption{Computational costs (in seconds) for one-way FMANOVA.}
	\vspace{1mm}
	\scalebox{0.9}{
	\begin{tabular*}{\textwidth}{@{\extracolsep{\fill}}cccccccc@{\extracolsep{\fill}}}
		\toprule
		W, LH, P, R   & $T_{\FLH}^\N$, $T_{\FLH}^\B$ & $T_{\ZZC}^\N$, $T_{\ZZC}^\B$   &$T_{\QFZ}$ &$T_{\QFZ}^{\max}$ & $T_{\Z}$ & $T_{\MFW}$, $T_{\MFLH}$,$T_{\MFP}$  \\ \midrule
		1753.91  & 50.02  &0.15   &42.68   &26.99   & 0.24  & 29.15\\
		\bottomrule
	\end{tabular*}}
	\label{real.tab1}
\end{table}

Since the heteroscedastic one-way FMANOVA is highly significant, we further apply the tests under consideration to some contrast tests to determine whether any two of the three regions share the same underlying group mean spectra. The results of these contrast tests are presented in Table~\ref{real.tab2}. All the tests produced consistent results for the contrast tests ``Greece vs. Italy" and ``Greece vs. Spain", indicating that the mean spectra from Greece differ significantly from those of Italy and Spain. However, the tests under consideration provided conflicting results regarding whether the mean spectra from Italy and Spain are significantly different. Most of the tests concluded that the mean spectra are not significantly different at the 5\% significance level, with only  $T_{\QFZ}$ and $T_{\QFZ}^{\max}$ reported opposite results. Notably, the $p$-value of $T_{\QFZ}$ was only slightly below the 5\% threshold.  Moreover, in \cite{soh2023regularised}'s study, Italy and Spain were grouped together under the category ``non-Greece",  supporting the hypothesis that their mean spectra are not significantly different.
\begin{table}[!h]
		\centering
	\setlength{\tabcolsep}{1pt} 
	\caption{$P$-values for some contrast tests  for the real dataset.}
	\vspace{1mm}
	\scalebox{0.9}{
	\begin{tabular*}{\textwidth}{@{\extracolsep{\fill}}lccccccc@{\extracolsep{\fill}}} \toprule
		&W     & LH    & P     & R    & $T_{\FLH}^\N$  & $T_{\FLH}^\B$  & $T_{\ZZC}^\N$   \\ \midrule
		Greece vs. Italy &$<0.001$  &$<0.001$ &$<0.001$ &$<0.001$  &$<0.001$  &$<0.001$   &$<0.001$  \\
		Greece vs. Spain &$<0.001$  &$<0.001$ &$<0.001$ &$<0.001$  &$<0.001$  &$<0.001$   &$<0.001$  \\
		Italy vs. Spain &0.134 &0.136 &0.133 &0.273&0.152 &0.139& 0.080 \\
		\midrule
		& $T_{\ZZC}^\B$ &$T_{\QFZ}$ &$T_{\QFZ}^{\max}$ & $T_Z$ & $T_{\MFW}$  & $T_{\MFLH}$  & $T_{\MFP}$ \\
		\midrule
		Greece vs. Italy  &$<0.001$  &$<0.001$  &$<0.001$ &$<0.001$  &$<0.001$  &$<0.001$  &$<0.001$\\
		Greece vs. Spain  &$<0.001$  &$<0.001$  &$<0.001$ &$<0.001$  &$<0.001$  &$<0.001$  &$<0.001$\\
		Italy vs. Spain  &0.050 &0.044 &$<0.001$ &0.068 &0.214 &0.233 &0.195\\
		\bottomrule
	\end{tabular*}}
	\label{real.tab2}
\end{table}

\section{Concluding Remarks}\label{con.sec}
In this paper, we introduced three tests, MFW, MFLH, and MFP for addressing the general linear hypothesis testing problem for multivariate functional data. These tests are highly adaptable and can be applied to a variety of scenarios by specifying appropriate contrast matrices. To illustrate their versatility, we considered applications such as the two-sample problem, one-way FMANOVA, and specific linear hypotheses. The methodology is based on two symmetric, nonnegative-definite matrices that capture the variation due to the hypothesis and error, with their distributions approximated by Wishart distributions. By aligning the mean and total variances of these matrices with their Wishart-approximations, we derived the approximate degrees of freedom that are estimated using a U-statistics-based approach. The resulting tests are affine-invariant and robust to deviations from Gaussianity and heteroscedasticity, showing strong performance even with small sample sizes.

Simulation studies revealed that, under equal covariance functions across groups, the proposed tests achieve comparable size control to the permutation-based tests of \cite{gorecki2017multivariate} and the bootstrap-based test of \cite{qiu2024tests}. Importantly, under heteroscedasticity, the proposed tests demonstrated superior performance, underscoring their robustness and utility in diverse testing scenarios. Furthermore, as shown in the real data analysis, the proposed tests are significantly faster computationally compared to the permutation-based and bootstrap-based alternatives. However, as indicated in Table~\ref{power.tab1}, the proposed tests are less powerful than $T_{\QFZ}^{\max}$ in \cite{qiu2024tests} when the functional data exhibit high or moderate correlations. Future work to enhance the power of the proposed tests is warranted and represents an interesting avenue for further research. In addition, although the use of the Wishart-approximation is not grounded in a formal theoretical foundation, the simulation results clearly indicate its appropriateness. Further investigation into this theoretical aspect is planned.

\section*{Acknowledgements}
The author is thankful to Dr. Zhu Ying and Dr. Soh Chin Gi for providing their dataset, which is used in Section~\ref{real.sec}, and to Dr. Zhiping Qiu for providing the code from \cite{Qiu2021,qiu2024tests}.

\appendix
\section{Technical Proofs}\label{proof.sec}
\setcounter{equation}{0}
\global\long\def\theequation{A.\arabic{equation}}
\begin{lemma}\label{lem1}
	Suppose that at each time point $t\in\calT$, $\bX(t)=[\bx_1(t),\ldots,\bx_m(t)]^\T$ is an $m\times p$ random matrix whose columns $\bx_i(t)\sim\SP_p(\bm{0},\bGamma_i)$ and are independent for $i=1,\ldots,m$.   Let $\bA=(a_{ij})$ be an $m\times m$  symmetric matrix, and  define $\bQ=\int_{\calT}\bX(t)^\T\bA\bX(t)dt$. The mean and total variation of $\bQ$ are given by
	\[
	 \E(\bQ)=\sum_{i=1}^m a_{ii}\bSigma_i\; \mbox{and}\; V(\bQ)=\sum_{i=1}^ma_{ii}^2\calK_4(\bx_i)+\sum_{i_1=1}^m\sum_{i_2=1}^ma_{i_1i_2}^2(I_{i_1i_2}+T_{i_1i_2}),
	\]
	 where  $\calK_4(\bx_i)=\int_{\calT^2}\E[\bx_{i}(s)^\T\bx_i(t)\bx_{i}(s)^\T\bx_{i}(t)]dsdt-\tr(\bSigma_i^2)-I_{ii}-T_{ii},i=1,\ldots,m$, $\bSigma_i=\int_{\calT}\bGamma_i(t,t) dt$, $I_{i_1i_2}=\int_{\calT^2}\tr[\bGamma_{i_1}(s,t)]\tr[\bGamma_{i_2}(s,t)]dsdt$ and $T_{i_1i_2}=\tr(\bGamma_{i_1}\bGamma_{i_2}),i_1,i_2=1,\ldots,m$.
\end{lemma}

\noindent{\bf Proof of Lemma~\ref{lem1}.}
We can further write $\bQ=\sum_{i_1=1}^m\sum_{i_2=1}^m a_{i_1i_2}\int_{\calT}\bx_{i_1}(t)\bx_{i_2}(t)^\T dt$. Since $\bx_i(t),i=1,\ldots,m$ are independently distributed, it is  easy to find 
	\[
	\E(\bQ)
	=\E\Big[\sum_{i=1}^m a_{ii}\int_{\calT}\bx_i(t)\bx_i(t)^\T dt\Big]=\sum_{i=1}^m a_{ii}\int_{\calT}\bGamma_i(t,t)dt=\sum_{i=1}^m a_{ii}\bSigma_i.
	\]
	Next, to find $V(\bQ)$, we have
	\[
	\begin{array}{rcl}
		V(\bQ)&=&\sum_{h=1}^p\sum_{\ell=1}^p\Var[\int_{\calT}\sum_{i_1=1}^m\sum_{i_2=1}^ma_{i_1i_2}x_{i_1h}(t)x_{i_2\ell}(t)dt]\\
		&=&\sum_{h=1}^p\sum_{\ell=1}^p\E[\int_{\calT}\sum_{i_1=1}^m\sum_{i_2=1}^ma_{i_1i_2}x_{i_1h}(t)x_{i_2\ell}(t)dt]^2\\
		&-&\sum_{h=1}^p\sum_{\ell=1}^p\E^2[\int_{\calT}\sum_{i_1=1}^m\sum_{i_2=1}^ma_{i_1i_2}x_{i_1h}(t)x_{i_2\ell}(t)dt]\\
		&=&V_1-V_2.
	\end{array}
	\]
	It follows that
	\[
	\begin{array}{rcl}
		V_1&=&\sum_{h=1}^p\sum_{\ell=1}^p\E[\int_{\calT^2}\sum_{i_1=1}^m\sum_{i_2=1}^m\sum_{i_3=1}^m\sum_{i_4=1}^ma_{i_1i_2}a_{i_3i_4}x_{i_1h}(s)x_{i_2\ell}(s)x_{i_3h}(t)x_{i_4\ell}(t)dsdt]\\
		&=&\sum_{i=1}^ma^2_{ii}\int_{\calT^2}\E[\bx_{i}(s)^\T\bx_i(t)\bx_{i}(s)^\T\bx_{i}(t)]dsdt+\sum_{i_1\neq i_2}\int_{\calT^2}a_{i_1i_1}a_{i_2i_2}\tr[\bGamma_{i_1}(s,s)\bGamma_{i_2}(t,t)]dsdt\\
		&+&\sum_{i_1\neq i_2}a_{i_1i_2}^2\int_{\calT^2}\tr[\bGamma_{i_1}(s,t)]\tr[\bGamma_{i_2}(s,t)]dsdt+\int_{\calT^2}\sum_{i_1\neq i_2}a_{i_1i_2}^2\tr[\bGamma_{i_1}(s,t)\bGamma_{i_2}(s,t)]dsdt\\
		&=&\sum_{i=1}^m a^2_{ii}\int_{\calT^2}\E[\bx_{i}(s)^\T\bx_i(t)\bx_{i}(s)^\T\bx_{i}(t)]dsdt+\sum_{i_1\neq i_2}a_{i_1i_1}a_{i_2i_2}\tr(\bSigma_{i_1}\bSigma_{i_2})\\
		&+&\sum_{i_1\neq i_2}a_{i_1i_2}^2I_{i_1i_2}+\sum_{i_1\neq i_2}a_{i_1i_2}^2T_{i_1i_2},
	\end{array}
	\]
	and 
	\[
	\begin{array}{rcl}
		V_2  &=&  \sum_{h=1}^p\sum_{\ell=1}^p\E^2[\int_{\calT}\sum_{i=1}^ma_{ii}x_{ih}(t)x_{i\ell}(t)dt]= \sum_{h=1}^p\sum_{\ell=1}^p[\int_{\calT}\sum_{i=1}^ma_{ii}\gamma_{i,h\ell}(t,t)dt]^2\\
		&=&\sum_{h=1}^p\sum_{\ell=1}^p\int_{\calT^2}\sum_{i_1=1}^m\sum_{i_2=1}^ma_{i_1i_1}a_{i_2i_2}\gamma_{i_1,h\ell}(s,s)\gamma_{i_2,h\ell}(t,t)dsdt\\
		&=&\int_{\calT^2}\sum_{i_1=1}^m\sum_{i_2=1}^ma_{i_1i_1}a_{i_2i_2}\tr[\bGamma_{i_1}(s,s)\bGamma_{i_2}(t,t)]dsdt\\
		&=&\sum_{i_1=1}^m\sum_{i_2=1}^ma_{i_1i_1}a_{i_2i_2}\tr(\bSigma_{i_1}\bSigma_{i_2}).
	\end{array}
	\]
	Therefore,
	\[
	V(\bQ)
	=\sum_{i=1}^ma_{ii}^2\calK_4(\bx_i)+\sum_{i_1=1}^m\sum_{i_2=1}^ma_{i_1i_2}^2(I_{i_1i_2}+T_{i_1i_2}),
	\]
	where $\calK_4(\bx_i)=\int_{\calT^2}\E[\bx_{i}(s)^\T\bx_i(t)\bx_{i}(s)^\T\bx_{i}(t)]dsdt-\tr(\bSigma_i^2)-I_{ii}-T_{ii},i=1,\ldots,m$. 

\begin{lemma}\label{Ustat.lem}
	Given the $k$ functional samples  (\ref{ksamp.sec2}), let $\bSigma_i=\int_{\calT}\bGamma_i(t,t)dt,i=1,\ldots,k$, \\ $I_{i_1i_2}=\int_{\calT^2}\tr[\bGamma_{i_1}(s,t)]\tr[\bGamma_{i_2}(s,t)]dsdt$, and $T_{i_1i_2}=\tr(\bGamma_{i_1}\bGamma_{i_2})=\int_{\calT^2}\tr[\bGamma_{i_1}(s,t)\bGamma_{i_2}(s,t)]dsdt$, $i_1,i_2=1,\ldots,k$. When $i_1\neq i_2$, the unbiased estimators of $I_{i_1i_2}$ and $T_{i_1i_2}$ are given by
	\begin{equation}\label{estuneq.equ}
		\hI_{i_1i_2} = \int_{\calT^2}\tr[\hbGamma_{i_1}(s,t)]\tr[\hbGamma_{i_2}(s,t)]dsdt,\mbox{ and } \hT_{i_1i_2} = \tr(\hbGamma_{i_1}\hbGamma_{i_2}),i_1\neq i_2,
	\end{equation}
	respectively, where $\hbGamma_i(s,t),i=1,\ldots,k$ is the sample matrix of group covariance functions given in (\ref{kmeancov.sec2}). In addition, the unbiased estimators of $I_{ii}, T_{ii}$ and $\tr(\bSigma_i^{2}),i=1,\ldots,k$ are 
	\begin{equation}\label{esteq.equ}
		\begin{array}{rcl}
			\hI_{ii}&=&\frac{1}{n_i(n_i-1)}\sum_{j_1\neq j_2}\int_{\calT^2}\by_{ij_1}(t)^\T\by_{ij_1}(s)\by_{ij_2}(t)^\T\by_{ij_2}(s)dsdt\\
			&-&\frac{2}{n_i(n_i-1)(n_i-2)}\sum_{j_1,j_2,j_3}^*\int_{\calT^2}\by_{ij_1}(t)^\T\by_{ij_1}(s)\by_{ij_2}(t)^\T\by_{ij_3}(s)dsdt\\
			&+&\frac{1}{n_i(n_i-1)(n_i-2)(n_i-3)}\sum_{j_1,j_2,j_3,j_4}^*\int_{\calT^2}\by_{ij_1}(t)^\T\by_{ij_2}(s)\by_{ij_3}(t)^\T\by_{ij_4}(s)dsdt,\\
			\hT_{ii}&=&\frac{1}{n_i(n_i-1)}\sum_{j_1\neq j_2}\int_{\calT^2}\by_{ij_1}(t)^\T\by_{ij_2}(s)\by_{ij_2}(t)^\T\by_{ij_1}(s)dsdt\\
			&-&\frac{2}{n_i(n_i-1)(n_i-2)}\sum_{j_1,j_2,j_3}^*\int_{\calT^2}\by_{ij_1}(t)^\T\by_{ij_2}(s)\by_{ij_3}(t)^\T\by_{ij_1}(s)dsdt\\
			&+&\frac{1}{n_i(n_i-1)(n_i-2)(n_i-3)}\sum_{j_1,j_2,j_3,j_4}^*\int_{\calT^2}\by_{ij_2}(t)^\T\by_{ij_3}(s)\by_{ij_4}(t)^\T\by_{ij_1}(s)dsdt,\mbox{ and }\\
			\widehat{\tr(\bSigma_i^{2})}  &=&\frac{1}{n_i(n_i-1)}\sum_{j_1\neq j_2}\int_{\calT^2}\by_{ij_1}(s)^\T\by_{ij_2}(t)\by_{ij_2}(t)^\T\by_{ij_1}(s)dsdt\\
			&-&\frac{2}{n_i(n_i-1)(n_i-2)}\sum_{j_1,j_2,j_3}^*\int_{\calT^2}\by_{ij_1}(s)^\T\by_{ij_2}(t)\by_{ij_3}(t)^\T\by_{ij_1}(s)dsdt\\
			&+&\frac{1}{n_i(n_i-1)(n_i-2)(n_i-3)}\sum_{j_1,j_2,j_3,j_4}^*\int_{\calT^2}\by_{ij_2}(s)^\T\by_{ij_3}(t)\by_{ij_4}(t)^\T\by_{ij_1}(s)dsdt,
		\end{array}
	\end{equation}
	respectively.
\end{lemma}

\noindent{\bf Proof of Lemma~\ref{Ustat.lem}.}
Since $\hbGamma_i(s,t)$ (\ref{kmeancov.sec2}) is the unbiased estimator of $\bGamma_i(s,t),i=1,\ldots,k$, it is straightforward to show that $\E\{\tr[\hbGamma_i(s,t)]\}=\tr[\bGamma_i(s,t)],i=1,\ldots,k$. When $i_1\neq i_2$, we have
	\[
	\begin{array}{rcl}
		\E(\hI_{i_1i_2}) &=&\int_{\calT^2}\E\{\tr[\hbGamma_{i_1}(s,t)]\}\E\{\tr[\hbGamma_{i_2}(s,t)]\}dsdt=\int_{\calT^2}\tr[\bGamma_{i_1}(s,t)]\tr[\bGamma_{i_2}(s,t)]dsdt=I_{i_1i_2}, \mbox{ and }\\
		\E(\hT_{i_1i_2})&=&\int_{\calT^2}\tr\{\E[\hbGamma_{i_1}(s,t)]\E[\hbGamma_{i_2}(s,t)]\}dsdt=\tr(\bGamma_{i_1}\bGamma_{i_2})=T_{i_1i_2}. 
	\end{array}
	\]
	Thus, equation (\ref{estuneq.equ}) is obtained. Next, we focus on demonstrating the unbiasedness of the estimators in (\ref{esteq.equ}). The main ideas of constructing the unbiased estimators in (\ref{esteq.equ}) follow U-statistics-based estimate approach in \cite{li2012two}. Besides considering U-statistics of the form $[n_i(n_i-1)]^{-1}\sum_{j_1\neq j_2}\int_{\calT^2}\by_{ij_1}(s)^\T\by_{ij_1}(t)\by_{ij_2}(s)^\T\by_{ij_2}(t)dsdt$, to address the case where $\E[\by_{ij}(t)]\neq \bm{0}$, we subtract two additional U-statistics of orders three and four, respectively. This approach dates back to \cite{glasser1961unbiased,glasser1962estimators} and follows the method outlined in equation (2.1) of \cite{li2012two}. Firstly,
	\[
	\begin{array}{rcl}
		\E(\hI_{ii})&=&\frac{1}{n_i(n_i-1)}\sum_{j_1\neq j_2}\int_{\calT^2}\E[\by_{ij_1}(t)^\T\by_{ij_1}(s)]\E[\by_{ij_2}(t)^\T\by_{ij_2}(s)]dsdt\\
		&-&\frac{2}{n_i(n_i-1)(n_i-2)}\sum_{j_1,j_2,j_3}^*\int_{\calT^2}\E[\by_{ij_1}(t)^\T\by_{ij_1}(s)]\E[\by_{ij_2}(t)^\T]\E[\by_{ij_3}(s)]dsdt\\
		&+&\frac{1}{n_i(n_i-1)(n_i-2)(n_i-3)}\sum_{j_1,j_2,j_3,j_4}^*\int_{\calT^2}\E[\by_{ij_1}(t)^\T]\E[\by_{ij_2}(s)]\E[\by_{ij_3}(t)^\T]\E[\by_{ij_4}(s)]dsdt\\
		&=&\frac{1}{n_{i}(n_{i}-1)}\sum_{j_1\neq j_2}\int_{\calT^2}\{\tr[\bGamma_i(s,t)]+\bfeta_i(t)^\T\bfeta_i(s)\}\{\tr[\bGamma_i(s,t)]+\bfeta_i(t)^\T\bfeta_i(s)\}dsdt\\
		&-&\frac{2}{n_i(n_i-1)(n_i-2)}\sum_{j_1,j_2,j_3}^*\int_{\calT^2}\{\tr[\bGamma_i(s,t)]+\bfeta_i(t)^\T\bfeta_i(s)\}\bfeta_i(t)^\T\bfeta_i(s)dsdt\\
		&+&\frac{1}{n_i(n_i-1)(n_i-2)(n_i-3)}\sum_{j_1,j_2,j_3,j_4}^*\int_{\calT^2}\bfeta_i(t)^\T\bfeta_i(s)\bfeta_i(t)^\T\bfeta_i(s)dsdt\\
		&=&\int_{\calT^2}\tr^2[\bGamma_i(s,t)]dsdt =I_{ii}.
	\end{array}
	\]
	Similarly, we have
	\begin{equation*}
	\begin{array}{rcl}
		\E(\hT_{ii})&=&\frac{1}{n_i(n_i-1)}\sum_{j_1\neq j_2}\int_{\calT^2}\tr\{\E[\by_{ij_1}(s)\by_{ij_1}(t)^\T]\E[\by_{ij_2}(s)\by_{ij_2}(t)^\T]\}dsdt\\
		&-&\frac{2}{n_i(n_i-1)(n_i-2)}\sum_{j_1,j_2,j_3}^*\int_{\calT^2}\tr\{\E[\by_{ij_1}(s)\by_{ij_1}(t)^\T]\E[\by_{ij_2}(s)]\E[\by_{ij_3}(t)^\T]\}dsdt\\
		&+&\frac{1}{n_i(n_i-1)(n_i-2)(n_i-3)}\sum_{j_1,j_2,j_3,j_4}^*\int_{\calT^2}\E[\by_{ij_2}(t)^\T]\E[\by_{ij_3}(s)]\E[\by_{ij_4}(t)^\T]\E[\by_{ij_1}(s)]dsdt\\
		&=&\frac{1}{n_i(n_i-1)}\sum_{j_1\neq j_2}\int_{\calT^2}\tr\{[\bGamma_i(s,t)+\bfeta_i(s)\bfeta_i(t)^\T][\bGamma_i(s,t)+\bfeta_i(s)\bfeta_i(t)^\T]\}dsdt\\
		&-&\frac{2}{n_i(n_i-1)(n_i-2)}\sum_{j_1,j_2,j_3}^*\int_{\calT^2}\tr\{[\bGamma_i(s,t)+\bfeta_i(s)\bfeta_i(t)^\T]\bfeta_i(s)\bfeta_i(t)^\T\}dsdt\\
		&+&\frac{1}{n_i(n_i-1)(n_i-2)(n_i-3)}\sum_{j_1,j_2,j_3,j_4}^*\int_{\calT^2}\bfeta_i(t)^\T\bfeta_i(s)\bfeta_i(t)^\T\bfeta_i(s)dsdt\\
		&=&\int_{\calT^2}\tr[\bGamma_i(s,t)\bGamma_i(s,t)]dsdt=T_{ii}.
	\end{array}
	\end{equation*}
	Finally, the expectation of $\widehat{\tr(\bSigma_i^{2})}$ can be obtained as follows:
	\[
	\begin{array}{rcl}
		\E[\widehat{\tr(\bSigma_i^{2})}] 	&=&\frac{1}{n_i(n_i-1)}\sum_{j_1\neq j_2}\int_{\calT^2}\tr\{\E[\by_{ij_1}(s)\by_{ij_1}(s)^\T]\E[\by_{ij_2}(t)\by_{ij_2}(t)^\T]\}dsdt\\
		&-&\frac{2}{n_i(n_i-1)(n_i-2)}\sum_{j_1,j_2,j_3}^*\int_{\calT^2}\tr\{\E[\by_{ij_1}(s)\by_{ij_1}(s)^\T]\E[\by_{ij_2}(t)]\E[\by_{ij_3}(t)^\T]\}dsdt\\
		&+&\frac{1}{n_i(n_i-1)(n_i-2)(n_i-3)}\sum_{j_1,j_2,j_3,j_4}^*\int_{\calT^2}\E[\by_{ij_2}(s)^\T]\E[\by_{ij_3}(t)]\E[\by_{ij_4}(t)^\T]\E[\by_{ij_1}(s)]dsdt\\
		&=&\frac{1}{n_i(n_i-1)}\sum_{j_1\neq j_2}\int_{\calT^2}\tr\{[\bGamma_i(s,s)+\bfeta_i(s)\bfeta_i(s)^\T][\bGamma_i(t,t)+\bfeta_i(t)\bfeta_i(t)^\T]\}dsdt\\
		&-&\frac{2}{n_i(n_i-1)(n_i-2)}\sum_{j_1,j_2,j_3}^*\int_{\calT^2}\tr\{[\bGamma_i(s,s)+\bfeta_i(s)\bfeta_i(s)^\T]\bfeta_i(t)\bfeta_i(t)^\T\}dsdt\\
		&+&\frac{1}{n_i(n_i-1)(n_i-2)(n_i-3)}\sum_{j_1,j_2,j_3,j_4}^*\int_{\calT^2}\bfeta_i(s)^\T\bfeta_i(t)\bfeta_i(t)^\T\bfeta_i(s)dsdt\\
		&=&\int_{\calT^2}\tr[\bGamma_i(s,s)\bGamma_i(t,t)]dsdt=\tr(\bSigma_i^2).
	\end{array}
	\]
	The proof is complete.
\newline

\noindent{\bf Proof of Proposition~\ref{afftrans1.prop}.}
	Under the affine-transformation (\ref{afftrans.equ}), the associated vector of mean functions and matrix of covariance functions transform as $\bfeta_i^0(t)=\bA\bfeta_i(t)+\bb(t),i=1,\ldots,k$ and $\bGamma^0_i(s,t)=\bA\bGamma_i(s,t)\bA^\T,i=1,\ldots,k$, respectively. Consequently,  the GLHT problem (\ref{GLHTH0.sec2}) associated with the transformed data becomes $\H_0^0:\bC\bM^0(t)=\bC_0^0(t),t\in\calT \mbox{ vs } \H^0_1:\bC\bM^0(t)\neq\bC_0^0(t), \mbox{ for some }t\in\calT$, where $\bM^0(t)=[\bfeta_1^0(t),\ldots,\bfeta_k^0(t)]^\T$, and $\bC_0^0(t)=\bC_0(t)\bA^\T+\bC\bm{1}_k\bb(t)^\T$. Since $\barby_i^0(t)=\bA\barby_i(t)+\bb(t)$, we have $\bC\hbM^0(t)-\bC_0^0(t)=[\bC\hbM(t)-\bC_0(t)]\bA^\T$, where $\hbM^0(t)=[\barby_1^0(t),\ldots,\barby_k^0(t)]^\T$ which is an unbiased estimator of $\bM^0(t)$. It follows that $\bB_n^0=\int_{\calT}\bA[\bC\hbM(t)-\bC_0(t)]^\T(\bC\bD_n\bC^\T)^{-1}[\bC\hbM(t)-\bC_0(t)]\bA^\T dt=\bA\bB_n\bA^\T$. Similarly, we have $\bOmega_n^0=\bA\bOmega_n\bA^\T$ and $\bE_n^0=\bA\bE_n\bA^\T$.  Additionally,  $\tr[\bGamma^*_i(s,t)]=\tr[\bOmega_n^{-1}\bGamma_i(s,t)]$ and $\tr(\bGamma_{i_1}^*\bGamma_{i_2}^*)=\int_{\calT^2}\tr[\bOmega^{-1}\bGamma_{i_1}(s,t)\bOmega^{-1}\bGamma_{i_2}(s,t)]dsdt$ are invariant under the affine-transformation (\ref{afftrans.equ}).  Furthermore, it is easy to find that $\bm{H}_n$ is unaffected by the affine-transformation (\ref{afftrans.equ}). These results imply that the proposed test statistics $T_{\MFW},T_{\MFLH}$, and $T_{\MFP}$ (\ref{teststa.equ}) are affine-invariant. Moreover, the approximate degrees of freedom $d_B$ and $d_E$ in (\ref{dB.sec2}) and (\ref{dE.sec2}), along with their estimators  in (\ref{hdB.sec2}) and (\ref{hdE.sec2}), are also invariant.     The proof is complete. 
\newline

\noindent{\bf Proof of Proposition~\ref{afftrans2.prop}.}
	To demonstrate that the proposed tests $T_{\MFW}$, $T_{\MFLH}$, and $T_{\MFP}$ (\ref{teststa.equ}) are invariant under the transformation specified in (\ref{lineartrans.equ}), it is sufficient to establish that $\bB_n$, $\bE_n$, $d_B$, and $d_E$ remain invariant under the same transformation.  	Specifically, we have $\bP\bC\hbM(t)-\bP\bC_0(t)=\bP[\bC\hbM(t)-\bC_0(t)]$, and $(\bP\bC\bD_n\bC^\T\bP^\T)^{-1}=(\bP^{-1})^\T(\bC\bD_n\bC^\T)^{-1}\bP^{-1}$. It follows that
	\[
	\bB_n\to \int_{\calT}[\bC\hbM(t)-\bC_0(t)]^\T\bP^\T(\bP^{-1})^\T(\bC\bD_n\bC^\T)^{-1}\bP^{-1}\bP[\bC\hbM(t)-\bC_0(t)] dt=\bB_n,
	\] and $\bm{H}_n\to \bC^\T\bP^\T(\bP^{-1})^\T(\bC\bD_n\bC^\T)^{-1}\bP^{-1}\bP\bC=\bm{H}_n$, both of which remain invariant under the transformation in  (\ref{lineartrans.equ}). Then the invariance of $\bE_n$, $d_B$, and $d_E$ follows immediately. 
\newline

\noindent{\bf Proof of Theorem~\ref{dBdEsol.thm}.}
	We first find $d_B$ so that $\tbB_n\sim W_p(d_B,\bI_p/d_B)$ approximately via matching the total variations of $\tbB_n$ and $\tbB$ where $\tbB\sim W_p(d_B,\bI_p/d_B)$. It can be shown that the total variation of $\tbB$
	is given by $V(\tbB)=d_B(p/d_B^2+p^2/d_B^2)=p(p+1)/d_B$.  Then we aim to find the total variation of $\tbB_n=\int_{\calT}\bOmega_n^{-1/2}\bX(t)^\T\bm{H}_n\bX(t)\bOmega_n^{-1/2}dt$. Let $\bx^*_{ij}(t)=\bOmega_n^{-1/2}\bx_{ij}(t),$ so that $\bx^*_{ij}(t)\iidsim \SP_p(\bm{0},\bGamma^*_i), j=1,\ldots,n_{i};\;i=1,\ldots,k$, where $\bGamma^*_i(s,t)=\bOmega_n^{-1/2}\bGamma_{i}(s,t)\bOmega_n^{-1/2},i=1,\ldots,k$. Let $\barbx_i^*(t)=\bOmega_n^{-1/2}\barbx_i(t),i=1,\ldots,k$ and  $\bar{\bX}^*(t)=[\barbx_1^*(t),\ldots,\barbx_k^*(t)]^\T$. As a result, we can express $\tbB_n=\int_{\calT}\bar{\bX}^*(t)^{\T}\bm{H}_n\bar{\bX}^*(t)dt$. Since $\E[\barbx_i^*(t)]=\bm{0}$ and $\Cov[\barbx_i^*(s),\barbx_i^*(t)]=\bGamma_i^*(s,t)/n_i,i=1,\ldots,k$, applying Lemma~\ref{lem1} yields
	\[
	V(\tbB_n)=\sum_{i=1}^kh_{ii}^2\calK_4(\barbx_i^*)+\sum_{i_1=1}^k\sum_{i_2=1}^k\frac{h_{i_1i_2}^2}{n_{i_1} n_{i_2}}(I_{i_1i_2}^*+T_{i_1i_2}^*),
	\]
	where $\calK_4(\barbx^*_i)=\int_{\calT^2}\E[\barbx^*_{i}(s)^\T\barbx^*_i(t)\barbx^*_{i}(s)^\T\barbx^*_{i}(t)dsdt]-\tr(\bSigma^{*2}_i)/n_i^2-I_{ii}^*/n_i^2-T_{ii}^*/n_i^2$, with $\bSigma_i^*=\int_{\calT}\bGamma^*_i(t,t)dt$, $I_{i_1i_2}^*=\int_{\calT^2}\tr[\bGamma_{i_1}^*(s,t)]\tr[\bGamma_{i_2}^*(s,t)]dsdt$, and $T_{i_1i_2}^*=\tr(\bGamma_{i_1}^*\bGamma_{i_2}^*)$.  
	
	Further, since $\int_{\calT^2}\E[\barbx^*_{i}(s)^\T\barbx^*_i(t)\barbx^*_{i}(s)^\T\barbx^*_{i}(t)dsdt] =\quad$\hfill   $n_i^{-3}\int_{\calT^2}\E[\bx^*_{i1}(s)^\T\bx^*_{i1}(t)\bx^*_{i1}(s)^\T\bx^*_{i1}(t)]dsdt+n_{i}^{-3}(n_{i}-1)[\tr(\bSigma^{*2}_i)+I_{ii}^*+T_{ii}^*]$, it follows that $\calK_4(\barbx^*_i)=\calK_4(\bx^*_{i1})/n_{i}^3$.
	Therefore, 
	\[
	V(\tbB_n)=\sum_{i=1}^k\frac{h_{ii}^2}{n_i^3}\calK_4(\bx^*_i)+\sum_{i_1=1}^k\sum_{i_2=1}^k\frac{h_{i_1i_2}^2}{n_{i_1} n_{i_2}}(I_{i_1i_2}^*+T_{i_1i_2}^*),
	\]
	where  $\calK_4(\bx^*_{i1})=\int_{\calT^2}\E[\bx^*_{i1}(s)^\T\bx^*_{i1}(t)\bx^*_{i1}(s)^\T\bx^*_{i1}(t)]dsdt-\tr(\bSigma^{*2}_i)-I_{ii}^*-T_{ii}^*,i=1,\ldots,k$. 
	
	By matching $V(\tbB_n)$ and $V(\tbB)$, equation (\ref{dB.sec2}) is derived. 
	
	
	Next, we proceed to find $d_E$ by matching $V(\tbE_n)$ and $V(\tbE)$. Similarly, it is straightforward to determine $V(\tbE)=p(p+1)/d_E$.  Thus, our task reduces to determining $V(\tbE_n)$. We can further express $\tbE_n$ as
	$\tbE_n=\sum_{i=1}^kh_{ii}\hbSigma_i^*/n_i$, where $\hbSigma_i^*=\int_{\calT}\hbGamma^*_i(t,t)dt$ with $\hbGamma^*_i(t,t)=(n_i-1)^{-1}\sum_{j=1}^{n_i}[\bx_{ij}^*(t)-\barbx_i^*(t)][\bx_{ij}^*(t)-\barbx_i^*(t)]^\T,i=1,\ldots,k$. It follows that 
	\[
	\E(\tbE_n)=\sum_{i=1}^kh_{ii}\E(\hbSigma^*_i)/n_i,\;\mbox{ and }\;V(\tbE_n)=\sum_{i=1}^kh^2_{ii}V(\hbSigma^*_i)/n_i^2.
	\]
	As a result, we aim to study $\hbSigma^*_i,i=1,\ldots,k$ first.  We can write 
	$\hbSigma_i^*=(n_i-1)^{-1}\int_{\calT}{\bX^*_i(t)}^{\T}\bP\bX_i^{*}(t) dt$,  
	where $\bX^*_i(t)=[\bx^*_{i1}(t),\ldots,\bx^*_{i n_i}(t)]^\T$ and $\bP=\bI_{n_i}-\bJ_{n_i}/n_i$ with $\bI_{n}$ denoting the usual $n \times n$ identity matrix and $\bJ_n$ denoting the usual  $n \times n$ matrix of ones. Then by Lemma~\ref{lem1}, we have $\E(\hbSigma_i^*)=(n_i-1)^{-1}\sum_{j=1}^{n_i}p_{jj}\bSigma^*_i=\bSigma^*_i$, provided by $p_{jj}=1-1/n_i$ for all $j=1,\ldots,n_i$. It follows that $\E(\tbE_n)=\sum_{i=1}^kh_{ii}\bSigma^*_i/n_i$. Furthermore,
	\[
	\begin{split}
		V(\hbSigma_i^*)&=(n_i-1)^{-2}\sum_{j=1}^{n_i}p_{jj}^2\calK_4(\bx_{i1}^*)+(n_i-1)^{-2}\sum_{j_1=1}^{n_i}\sum_{j_2=1}^{n_i}p_{j_1j_2}^2(I_{ii}^*+T_{ii}^*)\\
		&=n_i^{-1}\calK_4(\bx^*_{i 1})+(n_i-1)^{-1}(I_{ii}^*+T_{ii}^*).
	\end{split}
	\]
	It follows that 
	\[
	V(\tbE_n)=\sum_{i=1}^k\frac{h_{ii}^2}{n_i^3}\calK_4(\bx^*_{i1})+\sum_{i=1}^k\frac{h_{ii}^2}{n_{i}^2(n_i-1)}(I_{ii}^*+T_{ii}^*).
	\]
	By matching $V(\tbE_n)$ and $V(\tbE)$, equation (\ref{dE.sec2}) is derived.

\bibliographystyle{apalike}      
\bibliography{KSBFMFDref}

\end{document}